\documentclass[
aps,
prx,
showpacs,
preprintnumbers,
twocolumn,
superscriptaddress,
10pt,
]{revtex4-2}

\usepackage{packages}

\makeatletter
\@ifundefined{ifmainmode}{\newif\ifmainmode\mainmodetrue}{}%
\@ifundefined{ifsuppmode}{\newif\ifsuppmode\suppmodetrue}{}%
\@ifundefined{ifcombinedmode}{\newif\ifcombinedmode\combinedmodetrue}{}%
\makeatother

\newcommand{\id}{\mathds{1}}

\newcommand{\mpipks}{Max Planck Institute for the Physics of Complex Systems, N\"othnitzer Str.\ 38, 01187 Dresden, Germany}

\begin{document}

\author{Giovanni Cemin}
\email{gcemin@pks.mpg.de}
\affiliation{\mpipks}
\author{Marin Bukov}
\affiliation{\mpipks}
\author{Andrea Solfanelli}
\affiliation{\mpipks}

\clearpage
\title{Error correction and quantum information spreading on sparse random graphs}

\ifmainmode

\begin{abstract}
    Quantum error-correcting codes on graph geometries beyond regular euclidean lattices have emerged as a promising route toward robust quantum information processing.
    A major challenge in quantum error correction is the trade-off between favorable properties of codes constructed in high dimensionality and the need for locality in practical implementations on near-term hardware.
    Sparse intermediate geometries may combine the advantages of both, but how their structure governs quantum information dynamics and code performance remains unclear.
    Here we show that the spectral dimension $d_s$, defined from the low-energy spectrum of the graph Laplacian, is the universal parameter connecting these properties.
    Using Floquet random Clifford circuits on two distinct sparse graph architectures with tunable spectral dimension, we find that $d_s$ governs both quantum information spreading and the scaling of the resulting quantum error-correcting code distance.
    We identify a geometry-induced entanglement phase transition from localized area-law to ergodic volume-law, whose universality is determined by $d_s$ and exhibits long-range percolation criticality.
    In the ergodic phase, the circuit acts as a CSS code encoder whose code distance is set by the competition between two logical sectors --- the cost of severing the graph and the length of its shortest winding cycle --- which connectivity pushes in opposite directions. The spectral dimension organizes this trade-off, highlighting a maximum at intermediate $d_s$ retaining sparse connectivity and physical realizability.
\end{abstract}

\maketitle

\stoptoc

\section{Introduction}
Traditionally, the interaction pattern of quantum many-body systems has been dictated by the Euclidean geometry of the experimental platform. Condensed-matter systems are therefore commonly described by local Hamiltonians whose connectivity reflects the spatial arrangement of their microscopic constituents.
This paradigm has recently been overturned.
Advances in programmable quantum simulators based on atomic, molecular, and optical (AMO) platforms, including Rydberg atom arrays~\cite{browaeys2020many,ebadi2022quantum}, trapped ions~\cite{monroe2021programmable}, and cold atoms in optical cavities~\cite{ritsch2013cold},
now enable interactions with a highly tunable range and connectivity. As a consequence, the geometry governing quantum many-body dynamics is no longer necessarily the physical embedding space, but rather the interaction graph connecting the microscopic degrees of freedom.

This shift puts the interaction graph itself at the center of quantum many-body research. Its structure governs a variety of collective phenomena, from spin-glass phenomenology~\cite{gopalakrishnan2011frustration,kroeze2025directly} and self-organized criticality~\cite{helmrich2020signatures} to persistent collective coherence and anomalous Floquet heating
\cite{gorlach2025supercoherence,guo2026anomalousfloquet}.
The interaction graph properties are equally central to quantum technologies: quantum search algorithms~\cite{young2022tweezer,king2025optimal}, variational quantum algorithms~\cite{tabares2023variational},
quantum metrology~\cite{solfanelli2026robust}, all implicitly take advantage of the interaction graph connectivity.
More explicitly, nontrivial graph structure enables the realization of robust quantum memory~\cite{dennis2002topological} and quantum error correction (QEC) codes~\cite{knill1997theory}.
In particular, extending QEC codes beyond regular Euclidean lattices to more general graph geometries has proven to be a route toward good quantum codes~\cite{rakovszky2023physicsgoodldpc,rakovszky2024physicsgoodldpc}, i.e., codes capable of correcting errors of extensive size~\cite{breuckmann2021product,panteleev2022asymptotically,dinur2023good,leverrier2022quantum}.
These codes rely on expander-like graph geometries~\cite{sipse1996expander,hoory2006expander}, whose connectivity is not geometrically local, whereas Euclidean lattice architectures face fundamental trade-offs that limit their performance~\cite{brabyi2010tradeoffs}.
This motivates the study of intermediate graph geometries, such as sparse long-range random graphs and fractal structures, which partially retain the connectivity advantages of expanders while preserving the locality required for practical implementation~\cite{yang2025spatiallycoupled,berthusen2025toward,berthusen2024partialsyndrome,dai2025locality,williamson2024layer,baspin2026wirecodes,sriram2026diffusion}.

Between strictly local lattices and expander graphs lies a continuum of geometries, yet no single quantity organizes geometries along this continuum, nor does it indicate what that position implies for the quantum information dynamics it supports and for the performance of the corresponding QEC code.
Entanglement dynamics directly probes the geometry, distinguishing phases with localized or delocalized quantum information.
This directly relates to the performance of the circuit as an encoder: a circuit in the delocalized phase conceals information from local probes, which is the defining property of a QEC code~\cite {gullans2020dynamicalpurification}.

Recent years have witnessed remarkable progress in identifying dynamical entanglement transitions in monitored quantum circuits~\cite{li2019measurement,skinner2019measurement} and disordered systems~\cite{abanin2019colloquium}. In these settings, however, the interaction graph is held fixed while the transition is driven by tuning the measurement rate or the disorder strength.
We here raise the fundamental question: can the geometry of the interaction network alone determine the universal properties of entanglement dynamics? More specifically, is there a quantity characterizing the underlying graph geometry that universally controls the transition between localized and ergodic entanglement growth, and the performance of the corresponding QEC code, independently of microscopic details?

In this work, we tackle these questions by investigating the dynamics of quantum circuits on generic interaction graphs.
We identify the spectral dimension $d_s$, defined through the low-lying spectrum of the discrete graph Laplacian~\cite{mieghem2010graph}, as the fundamental quantity governing the universal properties of quantum information dynamics. The graph Laplacian is the generator of diffusion processes, and the spectral density of its low-lying eigenmodes governs the recurrence of random walks~\cite{burioni2005randomwalks}, the emergence of long-range order~\cite{cassi1992phase,cassi1996local,burioni1999inverse,burioni1999transience,bruno2001absence,maghrebi2017continuous,solfanelli2024universality}, and localization phenomena~\cite{levitov1989absence,levitov1990delocalization,rodriguez2000quantumdiffusion,mirlin1996transition,demoura1998delocalization,sarkar2026emergent} on the graph.
Here, we show that this framework, originally developed for the study of the statistical mechanics of classical complex networks \cite{albert2002staitiscal,dorogovtsev2008critical,villegas2023laplacian}, naturally extends to nonequilibrium quantum dynamics.
Within this framework, the spectral dimension provides a universal description of quantum information spreading and entanglement growth, as well as the performance of dynamically generated QEC codes.
Crucially, interaction graphs with tunable $d_s$ interpolate continuously between the two limits: expander-like graphs, characterized by effectively infinite spectral dimension, and local Euclidean geometries, where $d_s$ reduces to the physical dimension of the underlying platform.
We demonstrate that the transition between a localized area-law phase and an ergodic volume-law phase is entirely governed by the spectral properties of the interaction graph: $d_s$ uniquely determines the existence and universality class of the transition, establishing it as the effective dimension controlling entanglement criticality on complex interaction graphs.
Moreover, interpreting the circuit as an encoder, we show that $d_s$ determines the scaling of the code distance --- the size of the smallest undetectable error --- with system size, yielding increasingly favorable scalings as $d_s$ grows.

We benchmark this framework using Floquet random Clifford circuits~\cite{gottesman1997thesis,gottesman1998heisenberg,aaronson2004improved,chandran2015semiclassical} on two experimentally motivated classes of interaction graphs: the long-range (LR) graph, describing systems with algebraically decaying interactions~\cite{defenu2023longrange,defenu2024outofequilibrium}, and the power-of-two (PWR2) graph, a sparse hierarchical architecture recently proposed for quantum information processing~\cite{periwal2021programmable,gregory2020nonlocal}.
Their microscopic connectivity differs significantly in average degree --- i.e., the mean number of connections per site --- and their spectral properties are tunable, making them well suited to testing the universality of our predictions.
Both are also sparse: only $O(N)$ entangling operations are required per Floquet period on average, so that circuit depth per period scales linearly with system size.

A practical realization of our quantum circuits could be achieved in existing quantum processors based on neutral atoms in reconfigurable optical tweezer arrays~\cite{bluvstein2022quantum,bluvstein2024logical}, providing a realistic platform for exploring geometry-induced entanglement transitions.
Estimates based on current experimental parameters indicate that each step of the Floquet dynamics can be implemented with reliable fidelity up to $N \lesssim 10^{2}$.
For both families, we determine the complete nonequilibrium phase diagram and accurately extract the critical exponents from large-scale numerical simulations and finite-size scaling analysis.

Despite their distinct local structures, the spectral dimension of the interaction graph entirely controls the critical behavior: in both models, the entanglement transition belongs to the universality class of long-range percolation~\cite{gori2017onedimensional} at the corresponding spectral dimension.
First, we relate the transition to the probability that a node cuts the interaction graph, preventing quantum information from spreading.
Second, we show that the transition is a percolation transition in the evolving tableau description, where a system-spanning connected cluster signals the onset of volume-law entanglement.
These complementary pictures provide an intuitive geometric understanding
of the critical behavior while revealing its close connection to
classical percolation phenomena.

Finally, we study the circuit dynamics as an encoding circuit for a QEC code~\cite{brown2013short,gullans2020dynamicalpurification,darmawan2024lowdepth}.
In the ergodic phase, the circuit encodes a logical qubit whose code distance grows with system size.
We show that this distance is governed by two competing geometric quantities: the cost of disconnecting the graph and the length of its shortest nontrivial winding path. These define two logical sectors, associated respectively with cut-like and cycle-like operators.
Increasing connectivity enhances protection against cuts, but simultaneously creates shortcuts that reduce winding distances, resulting in a trade-off between the two error channels.
The spectral dimension organizes this trade-off by continuously interpolating between local Euclidean geometries, where winding paths are long but cuts are inexpensive, and highly connected or small-world graphs~\cite{watts1999smallworld}, where cuts become extensive but winding paths are short.
As a result, intermediate graph geometries can provide an optimal balance between the two sectors, yielding enhanced code-distance scaling while retaining sparse connectivity and physical realizability.

Together, these results connect the geometry of the interaction graph, the dynamics of entanglement generation, and the performance of dynamically generated QEC codes, identifying the spectral dimension as a tuning parameter for the design of robust quantum information architectures.

\begin{figure}[t]
    \centering
    \includegraphics{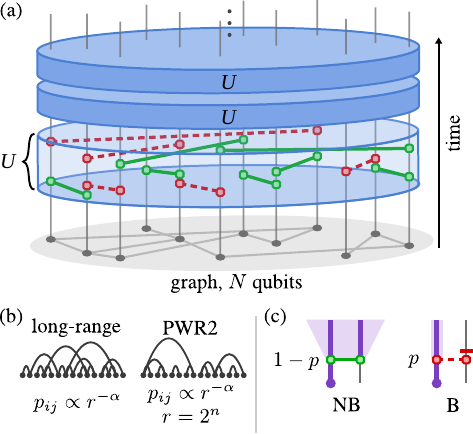}
    \caption{(a) Schematic of a Floquet random circuit on $N$ qubits living on a graph, generated by the stroboscopic application of $U$.
    The unitary $U$ is built by placing a two-qubit gate on every edge, non-blocking (NB, green solid line) with probability $1-p$ and blocking (B, red dashed line) with probability $p$.
    (b) Sample realizations of the long-range (LR) and power-of-two (PWR2) graphs embedded in 1D. Each pair of sites $(i,j)$ is connected with probability $p_{ij} \propto r^{-\alpha}$, where $r := |i-j|$ is the distance between sites $i$ and $j$; for PWR2, $r$ is restricted to powers of $2$.
    (c) The two gate types: the non-blocking (NB) gate spreads information across the edge it acts on, while the blocking (B) gate hosts a local integral of motion (LIOM) and prevents information from spreading across the bond.}
    \label{fig:model}
\end{figure}
\section{Quantum circuits on random graphs}
\label{sec:model}
Random circuits built from Haar-random two-qubit gates rapidly thermalize, leading to ballistic operator spreading and volume-law entanglement without any localization transition~\cite{nahum2017entanglementgrowth,nahum2018operatorspreading,fisher2023randomcircuits} unless additional dynamical constraints are imposed~\cite{vonKeyserkingk2018otocs,skinner2019measurement,khemani2018operatorspreading,li2019measurement}.
At the opposite extreme, static disordered Hamiltonians may exhibit ergodicity breaking through many-body localization, but their dynamics are complicated by the presence of slow modes associated with energy conservation and other global conserved quantities.
Floquet dynamics provides a clean setting where the periodic drive need not conserve energy or any other extensive quantity, and has been used extensively to investigate the stability of many-body localization~\cite{ponte2015manybody, bukov2015universal}.

We consider a class of Floquet random Clifford circuits on random interaction graphs.
Clifford circuits retain the competition between scrambling and localization while remaining efficiently simulable classically, giving access to the large system sizes required for finite-size scaling~\cite{gottesman1998heisenberg,aaronson2004improved}.

The dynamics is generated by the stroboscopic application of a unitary $U$ to $N$ qubits, so that the state after $t$ periods reads
\begin{equation}
    \ket{\psi (t)} = U^t \ket{\psi (0)}.
\end{equation}
$U$ is a multilayer circuit of two-qubit Clifford gates, specified by two independent ingredients: the interaction graph and the gates placed on its edges.
The former is a random graph on the $N$ qubits, in which each unordered pair $(i,j)$ is connected by an edge with probability $p_{ij}$ [Fig.~\ref{fig:model}(a)].
Every realized edge then hosts exactly one two-qubit gate, drawn independently as \textit{blocking} (B) with probability $p$ and \textit{non-blocking} (NB) with probability $1-p$~\cite{chandran2015semiclassical}.
The two gate types differ in whether they transmit quantum information.
A blocking gate $U_{\rm B}$ preserves a local integral of motion (LIOM), a local operator $O$ satisfying $[U_{\rm B}, O] = 0$, and therefore prevents information from propagating across its edge.
An NB gate instead scrambles local operators, allowing information and entanglement to spread [Fig.~\ref{fig:model}(c)].
The dynamics is thus fully determined by $p_{ij}$ and $p$ alone.

Our results depend only on the abstract B/NB structure, not on the specific Clifford gates realizing it (see App.~\ref{app:BNB_gates} for details).
We consider two representative implementations.
In the first,
\begin{equation}
    \begin{aligned}
    \mathrm{NB}&: ZI \mapsto ZZ, \; XI \mapsto ZY, \; IZ \mapsto XX, \; IX \mapsto YX, \\
    \mathrm{B}&: ZI \mapsto ZI, \; XI \mapsto XZ, \; IZ \mapsto IZ, \; IX \mapsto ZX
    \end{aligned}
    \label{eq:BNBdefinition}
\end{equation}
the blocking gate is the diagonal controlled-$Z$, which preserves every local $Z$-string, while the non-blocking gate scrambles operators across the bond~\cite{chandran2015semiclassical}.
In the second, B and NB are the identity and the CNOT, which makes the circuit a CSS code encoder~\cite{nielsen2010quantum}.

Floquet random Clifford circuits have been studied on Euclidean lattices, where the phenomenology depends sharply on dimension~\cite{farshi2023absence}: on a one-dimensional chain such circuits are always localized~\cite{farshi2022mixing}, whereas on two-dimensional lattices localized and ergodic phases are separated by a continuous transition governed by the percolation of ergodic clusters~\cite{chandran2015semiclassical}.
We here recover the Euclidean results as special cases, and we show how the resulting entanglement transition is universally controlled by the spectral properties of the graph.

\section{Entanglement transition and cutting probability}
\label{sec:entanglement_t_and_cut_prob}
\begin{figure}[t]
    \centering
    \includegraphics{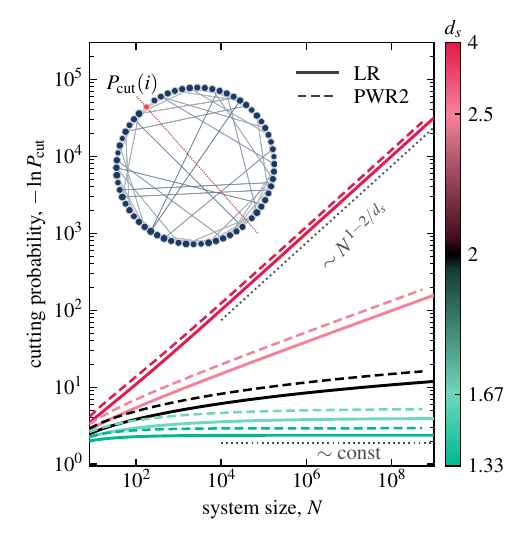}
    \caption{Cutting probability $P_{\rm cut}(i)$ across the threshold spectral dimension.
    Exact evaluation of Eqs.~\eqref{eq:long_range_lnPcut} and \eqref{eq:pwr2_lnPcut} at $p=0.5$, for the long-range (solid) and power-of-two (dashed) graphs at matched spectral dimension.
    The black line marks the critical value $d^*_s = 2$.
    For $d_s > d_s^*$, the cutting probability vanishes as a stretched exponential;
    for $d_s < d_s^*$ it saturates at a finite value and absolute cuts survive in the thermodynamic limit. Two microscopically different graph families with the same spectral dimension share the same exponent, differing only by a prefactor.
    Inset: an absolute cut at site $i$ requires every bond crossing it to carry a blocking gate.}
    \label{fig:p_cut}
\end{figure}
A localized phase can exist only if quantum information is prevented from propagating across the system.
On an arbitrary interaction graph, this requires blocking cuts: sites that disconnect the graph into independent regions.
We begin with the simplest such mechanism, an absolute cut characterized by a site that blocks every edge connecting two halves of the system; see Fig.~\ref{fig:p_cut}.
Although this condition is only sufficient, not necessary, for localization, it already yields an analytical criterion that reveals the central role of the graph geometry.

Site $i$ forms an absolute cut when every edge that straddles site $i$ or ends on it carries a blocking gate; its probability is
\begin{equation}
  P_{\rm cut}(i) = \prod_{\ell =0}^{i-1} \ \prod_{j=0}^{N-i}
        \left( 1 - \overline{p}_{i-\ell, \, i+j} \right),
    \label{eq:Pcut-product}
\end{equation}
with $\overline{p}_{i,j} = p_{i,j}(1-p)$; the full derivation is given in App.~\ref{app:cutting_p}.

If $P_{\rm cut}(i)>0$ as $N\to\infty$, a finite density of absolute cuts survives in the thermodynamic limit, quantum information cannot propagate across these cuts, and entanglement remains bounded by an area law.
The system is localized for any finite density of blocking gates, and the critical blocking probability is $p_c = 0$.
Conversely, if $P_{\rm cut}(i)\to 0$ as $N\to\infty$, isolated blocking sites disappear in the thermodynamic limit.
While this does not guarantee delocalization --- larger blocking clusters may still localize the dynamics --- it is a necessary condition for the emergence of an entanglement transition at $0<p<1$.

The asymptotic behavior of $P_{\rm cut}$ is determined by the geometry of the interaction graph, which enters through the graph Laplacian~\cite{mieghem2010graph}
\begin{equation}
    \mathbb{L}_{ij} = z_i\delta_{ij}-\mathbb{A}_{ij}\,,
\end{equation}
where $\mathbb{A}$ is the weighted adjacency matrix and $z_i = \sum_{j}\mathbb{A}_{ij}$ the weighted degree of node $i$.
$\mathbb{L}$ is Hermitian and admits a complete set of orthonormal eigenvectors $\{\mathbf{u}_n\}$ with non-negative eigenvalues $0=\lambda_0<\lambda_1\leq\dots\leq\lambda_{N-1}$, the zero mode corresponding to the uniform eigenvector $\mathbf{u}_0 = (1,\dots,1)/\sqrt{N}$.
The spectral dimension $d_s$ is defined by the scaling of the spectral gap $\lambda_1$ with system size~\cite{chung1994spectral},
\begin{equation}
    \lambda_1\sim N^{-2/d_s}
    \label{eq:spectral_gap_scaling},
\end{equation}
as $N\to\infty$.

Up to the edges joining site $i$ to its left neighbors, the edges entering Eq.~\eqref{eq:Pcut-product} are those crossing the bipartition at site $i$, and their total weight is a quadratic form of the graph Laplacian (App.~\ref{app:cutting_p}).
The variational (Rayleigh--Ritz) bound on this quadratic form~\cite{chung1994spectral,mieghem2010graph}, together with Eq.~\eqref{eq:spectral_gap_scaling}, bounds $P_{\rm cut}$ from above for any graph; under mild assumptions on the low-lying spectrum, a matching lower bound holds, and together they give
\begin{equation}
    P_{\rm cut}\sim
    \begin{cases}
        \mathrm{const.} & d_s<2,\\
        N^{-\theta(p)} & d_s=2,\\
        e^{-cN^{1-2/d_s}} & d_s>2.
    \end{cases}
    \label{eq:Pc_scaling}
\end{equation}
For $d_s < 2$, the graph is sparse enough that blocking cuts survive with finite probability as $N \to \infty$: quantum information is confined to finite regions, the system is localized, and the steady state obeys an area law.
At the marginal dimension $d_s = 2$, the cutting probability vanishes only algebraically, with a non-universal exponent $\theta(p)$ (App.~\ref{app:summary}).
For $d_s > 2$, blocking cuts disappear as a stretched exponential of the system size, so localization can no longer be sustained by isolated cuts and, if present, must emerge collectively, from extended clusters of blocking and non-blocking bonds.

This connects the entanglement transition to classical percolation: once absolute cuts disappear, localization is controlled by the percolation properties of the B/NB network.
Its critical behavior is expected to be governed by the spectral dimension rather than the microscopic graph topology~\cite{gori2017onedimensional}: for $d_s<2$ no extensive percolating cluster can form and the system remains localized, whereas for $d_s>2$ a finite threshold $p_c$ may exist below which a cluster spanning $O(N)$ sites emerges.
In the quantum circuit, such a cluster carries quantum information across the entire system, driving the transition from area-law to volume-law entanglement.

\section{Mapping to percolation within stabilizer formalism}
\label{sec:stab_formalism_percolation}
The stabilizer formalism (App.~\ref{app:stabilizer_formalism}) provides the connection to the percolation picture.
Every stabilizer state is equivalent, up to single-qubit Clifford gates, to a graph state,
\begin{equation}
    \ket{\psi} \;=\; \Big( \bigotimes_{i=1}^{N} C_i \Big) \ket{G_\psi} ,
    \qquad
    \ket{G_\psi} = \prod_{(i,j) \in E(G_\psi)} \mathrm{CZ}_{ij} \ket{+}^{\otimes N},
    \label{eq:graph_state}
\end{equation}
with $C_i \in \{ \id, H, S, \dots \}$ single-qubit Cliffords, obtained by bringing the tableau into canonical form.
Because the $C_i$ in Eq.~\eqref{eq:graph_state} are local, the entanglement structure of $\ket{\psi}$ is encoded in the graph $G_\psi$.
Notice the distinction between the two graphs: the \emph{interaction graph} of Sec.~\ref{sec:model} specifies where the gates act and is fixed by the realization; the \emph{entanglement graph} $G_\psi$ defined by Eq.~\eqref{eq:graph_state} is a property of the evolved state and is read off the tableau.

For a bipartition $A \cup \bar{A}$ the entropy is the $\mathbb{F}_2$ rank of the off-diagonal block of the adjacency matrix $\Theta = \Theta^T$ of $G_\psi$~\cite{hein2004multiparty,hein2006entanglement},
\begin{equation}
    S_A \;=\; \mathrm{rank}_{\mathbb{F}_2} \, \Theta_{A \bar{A}}.
    \label{eq:entropy_rank}
\end{equation}
If $A$ is a union of connected components of $G_\psi$, then $\Theta_{A\bar{A}} = 0$ and $S_A = 0$.
Decomposing
\begin{equation}
    G_\psi \;=\; \bigsqcup_{c} \, \mathcal{C}_c,
\end{equation}
the components $\mathcal{C}_c$ are the clusters of qubits scrambled together by the dynamics in the percolation picture.
They are obtained from $\Theta$ by a standard connected-components search in $O(N + |E|)$ time.

Blocking gates preserve a LIOM and do not generate edges of $G_\psi$ across the bond on which they act.
As a result, the components of $G_\psi$ are inherited from the connected components of the NB subgraph of the interaction graph, which they saturate once the dynamics has reached the steady state.

\subsection{Percolation}
The transition can be understood as a percolation transition of the entanglement clusters.
Let $\mathcal{C}_{\max}$ be the largest component and
\begin{equation}
    P_\infty \;=\; \frac{|\mathcal{C}_{\max}|}{N}
\end{equation}
the corresponding order parameter.
For $p > p_c$, blocking gates are dense enough that $G_\psi$ fragments into components of finite size;
any cut of $\Theta_{A\bar{A}}$ crosses at most $O(1)$ components, by Eq.~\eqref{eq:entropy_rank} the half-chain entropy saturates to $S_{N/2} = O(1)$, and the system fails to thermalize.
For $p < p_c$, a giant component spans the system, $P_\infty > 0$, the rank in Eq.~\eqref{eq:entropy_rank} grows with the size of the cut, and $S_{N/2}$ becomes extensive.
The critical point separating the two regimes is the percolation threshold of the entanglement graph $G_\psi$.

The universality class of the transition is expected to be determined by the geometry of the interaction graph, namely, by its spectral dimension $d_s$ (see Sec.~\ref{sec:entanglement_t_and_cut_prob}).
This allows the critical exponents extracted from the circuit to be compared directly with the universality class of classical long-range percolation~\cite{gori2017onedimensional} at the same spectral dimension.

\section{Experimentally motivated examples}
\begin{figure*}[t]
    \centering
    \includegraphics[width=\textwidth]{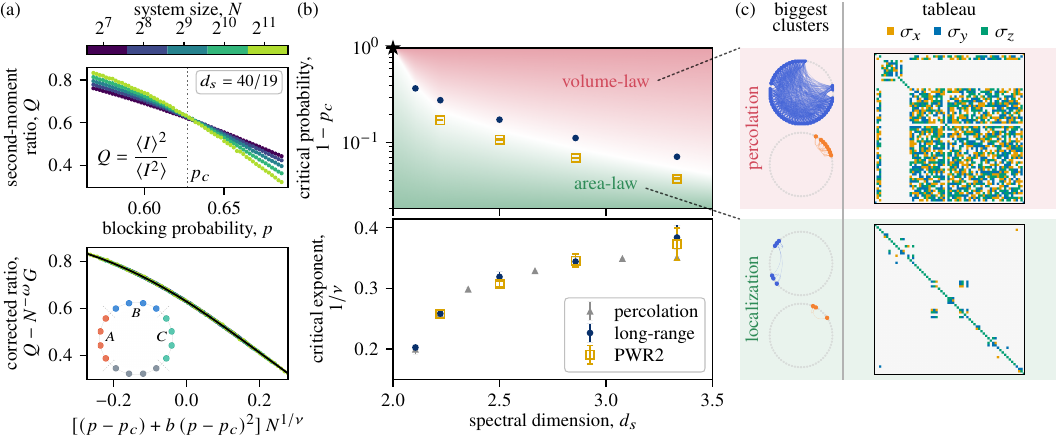}
    \caption{
    \textbf{Spectral dimension governs the criticality of Floquet Clifford circuits on random graphs.}
    (a) Finite-size scaling collapse of the second-moment ratio $Q = \langle I \rangle^2/\langle I^2\rangle$ of the quantum conditional mutual information $I(A:C\mid B)$, for the tripartition shown in the inset ($|A|=|B|=|C|=N/4$), at $\alpha = 1.95$ ($d_s = 40/19$).
    Top: raw $Q$ versus blocking probability $p$ for system sizes $N = 128$--$2048$.
    Bottom: corrected ratio $Q - N^{-\omega}G$ collapsed against the scaling variable $x$ of Eq.~\eqref{eq:fss_ansatz}.
    (b) Top: critical blocking probability $p_c$ versus spectral dimension $d_s$ for long-range and PWR2 graphs; the star marks the theoretical value $p_c = 0$ at the threshold spectral dimension $d_s^* = 2$.
    The shaded regions mark the volume-law and area-law phases; the gradient is a guide to the eye.
    Bottom: critical exponent $1/\nu$ versus $d_s$ for both graph families, compared with classical long-range percolation from \cite{gori2017onedimensional}.
    (c) Representative circuit realizations at $p < p_c$ (percolating, volume-law) and $p > p_c$ (localized, area-law): the two largest entanglement clusters (left) and the corresponding stabilizer tableau, colored by Pauli type $\sigma_x, \sigma_y, \sigma_z$ (right).
    }
    \label{fig:phase_diagram}
\end{figure*}
To validate our general theory, we consider two experimentally motivated families of graphs with different microscopic connectivity and continuously tunable spectral dimension: the long-range (LR) and the power-of-two (PWR2)~\cite{periwal2021programmable,bentsen2019treelike} graphs [see Fig.~\ref{fig:model}(b)].
Both are sparse random realizations of the long-range interactions arising naturally in several quantum simulation platforms, including trapped ions, Rydberg atom arrays, and cavity-mediated systems.

Neutral atoms in reconfigurable optical tweezer arrays~\cite{bluvstein2022quantum,bluvstein2024logical} offer a realistic route to implementing our circuits.
Local single-qubit rotations are readily achieved with microwave or Raman pulses, and high-fidelity controlled-phase gates are generated via the Rydberg blockade mechanism~\cite{xu2021fast,fromonteil2024hamilton,bluvstein2024logical}; combining local rotations with controlled-$Z$ operations synthesizes both the B and NB Clifford gates of our construction.
The interaction graph is then dynamically programmed by rearranging atomic positions through fast tweezer transport~\cite{bluvstein2022quantum} and ``faro-shuffle''-like swapping protocols~\cite{aldous1986shuffling,diaconis1983mathematics}, bringing different pairs of atoms within the blockade radius and entangling them in successive Floquet layers.
Repeating this sequence of parallel two-qubit gates and atom rearrangements realizes the desired graph connectivity~\cite{kuriyattil2025entangled}.

The sparse average connectivity of the LR and PWR2 graphs makes this implementation tractable: unlike all-to-all architectures, only $O(N)$ entangling operations are required on average per Floquet period, resulting in circuit depths per period that scale linearly with system size.
Estimates based on current experimental parameters indicate that the Floquet step can be implemented with reliable overall fidelity for system sizes up to $N\lesssim 10^{2}$ (App.~\ref{app:fidelityestimates}).

\subsection{Long-range graph}
\label{sec:lr_graph}
In this graph, every pair of sites separated by a distance $r = |i-j|$ on a $d$-dimensional lattice is connected with probability
\begin{equation}
    p_{i,j} = \frac{1}{r^{\alpha}}.
\end{equation}
This provides a sparse realization of systems with algebraically decaying interactions, which arise in a wide variety of programmable quantum simulators, including trapped ions, dipolar spin systems, and cavity-mediated atomic platforms~\cite{defenu2023longrange,bighin2024universal}.
It also models spatially correlated disorder ($p_{i,j}\neq p_ip_j$), in which the probability of long-range errors decays algebraically with distance~\cite{aharonov2006faultollerant,solfanelli2024stabilization,solfanelli2026robust}.
Beyond its experimental motivation, the LR graph is a paradigmatic setting for exploring universality~\cite{millan2021complex}: distinct universality classes are accessed continuously by tuning the interaction exponent $\alpha$, which makes it an ideal benchmark for our theoretical framework.

In the thermodynamic limit, the graph is self-averaging%
~\footnote{Deviations from the self-averaging description may arise near the crossover to short-range universality at $\alpha = d+2$~\cite{millan2021complex,liu2025twodimensional}, where logarithmic corrections are expected~\cite{brezin2014crossover}. Throughout this work, we restrict to the regime $1<\alpha<2$, which lies within the long-range universality regime and where corrections to self-averaging can be safely neglected.}, and its critical behavior is governed by the spectral dimension of the corresponding annealed graph, computed analytically in App.~\ref{app:ds_longrange}.
For a LR graph embedded in $d$-dimensional Euclidean space~\cite{millan2021complex,solfanelli2024universality},
\begin{align}
    d_s^{\rm LR} = \begin{cases}
        \infty &\alpha<d,\\
        2d/(\alpha-d) &d<\alpha<d+2,\\
        d &\alpha>d+2.
    \end{cases}\label{eq:dsLRgraph}
\end{align}
For sufficiently rapidly decaying interactions $(\alpha > d+2)$, the graph crosses over to the conventional nearest-neighbor universality class, recovering the Euclidean spectral dimension $d_s = d$.
Decreasing $\alpha$ enhances the effective connectivity and raises $d_s$; once $d_s$ exceeds the upper critical dimension~\footnote{For percolation the upper critical dimension is $d_c = 6$, above which the critical exponents take their mean-field values. Through Eq.~\eqref{eq:dsLRgraph}, $d_s > 6$ corresponds to $\alpha < 4d/3$, i.e.\ $\alpha < 4/3$ in $d=1$.}, the critical behavior continuously approaches the mean-field universality class.
For $\alpha<d$, the average coordination diverges and the graph becomes effectively mean-field, in the universality class of an Erd\H{o}s--R\'enyi graph~\cite{erdos1959random,erdos1960evolution}, leading to quantum information spreading and giving volume-law steady-state entanglement for any $p<1$.

The same spectral dimension determines the scaling of the cutting probability.
Translational invariance of the bond distribution allows us to evaluate $P_{\rm cut}$ analytically (App.~\ref{app:asympt}), yielding
\begin{align}
    -\ln P_{\rm cut}^{\rm LR} \sim
    \begin{cases}
        N^{2-\alpha}, & 0 < \alpha < 2, \\
        \ln N, & \alpha = 2, \\
        \mathrm{const.}, & \alpha > 2,
    \end{cases}
    \qquad N \gg 1,
\end{align}
where we set $d=1$ for simplicity.
For $\alpha>1$, substituting $\alpha$ for $d_s$ using Eq.~\eqref{eq:dsLRgraph} exactly reproduces the general scaling of Eq.~\eqref{eq:Pc_scaling}, and the criterion $\alpha=2$ translates into the threshold $d_s = 2$.
For $\alpha>2$ ($d_s^{\rm LR}<2$), a finite cutting probability guarantees localization at every finite density of B gates, and no entanglement transition survives in the thermodynamic limit.
The relevant regime is therefore the window $1<\alpha<2$, where isolated blocking cuts disappear while the graph remains sufficiently sparse to sustain a genuine geometry-induced entanglement transition.

\subsection{Power-of-two graph}
In this graph, interactions occur only between sites separated by dyadic distances~\cite{vladimirov1994p}: two sites at distance $r = |i-j|$ on a $d$-dimensional lattice are connected with probability
\begin{align}
    p_{i,j} =
    \begin{cases}
        r^{-\alpha}, & \text{for } r = 2^{n},\\
        0, & \text{otherwise},
    \end{cases}
\end{align}
with $n=0,1,\dots,\lfloor\log_2(N-1)\rfloor$.
Considerably sparser than the LR graph, this construction retains a continuously tunable spectral dimension, and therefore tests the universality of our predictions on a graph of entirely different microscopic connectivity.

Besides the reconfigurable Rydberg-array implementation~\cite{kuriyattil2025entangled} discussed above and in App.~\ref{app:fidelityestimates}, the PWR2 geometry can be realized in cavity-QED platforms with atoms trapped in optical tweezers and subject to spatially varying magnetic fields~\cite{periwal2021programmable}.
There, cavity-mediated spin-exchange interactions are selectively activated by magnetic-field gradients and periodic modulation of the drive, engineering interactions exclusively between sites separated by powers of two --- a direct analogue simulation of the PWR2 interaction graph.
The geometry has also attracted theoretical attention in the context of quantum information spreading~\cite{hashizume2021deterministic,hashizume2022tunable} and measurement-induced phase transitions~\cite{hashizume2022measurement}; at $\alpha = 0$ it gives rise to fast scrambling dynamics~\cite{bentsen2019treelike}.

The spectral dimension is obtained analytically (App.~\ref{app:ds_pwr2}),
\begin{align}
    d_s^{\rm PWR2} = \begin{cases}
        2/\alpha &0<\alpha<2,\\
        1 &\alpha>2.
    \end{cases}
    \label{eq:dsPWR2graph}
\end{align}
Despite the different microscopic connectivity, the PWR2 and LR graphs can realize the same universality class: in one dimension,
\begin{equation}
    d_s^{\rm PWR2} = d_s^{\rm LR}
\end{equation}
follows from
\begin{equation}
    \alpha_{\rm PWR2} = \alpha_{\rm LR}-1.
\end{equation}

The cutting probability likewise follows analytically (App.~\ref{app:asympt}),
\begin{align}
    -\ln P_{\rm cut}^{\rm PWR2}\sim
    \begin{cases}
        N^{1-\alpha}, & 0 < \alpha < 1, \\
        \ln N, & \alpha = 1, \\
        \mathrm{const.}, & \alpha > 1,
    \end{cases}
    \qquad N \gg 1,
\end{align}
again in agreement with the general scaling of Eq.~\eqref{eq:Pc_scaling} after substituting $\alpha$ for $d_s$ from Eq.~\eqref{eq:dsPWR2graph}.
The threshold $\alpha = 1$ therefore separates graphs with finite cutting probability ($d_s<2$) from those in which isolated cuts vanish in the thermodynamic limit ($d_s>2$), and the PWR2 graph supports a geometry-induced localization--delocalization transition only for $0<\alpha<1$.

\subsection{Numerical characterization of the entanglement transition}
For both graph families, we simulate the Floquet Clifford dynamics until reaching the nonequilibrium steady state.
We characterize the transition using the quantum conditional mutual information (QCMI)~\cite{otsuka2025qcmi},
\begin{equation}
    I(A:C|B) = S_{A \cup B} + S_{B \cup C} - S_B - S_{A\cup B \cup C} ,
\end{equation}
for the tripartition $|A| = |B| = |C| = N/4$ shown in Fig.~\ref{fig:phase_diagram}(a).
To mitigate finite-size effects, we cap the two-qubit gate range at $N/4$, consistent with the support of the observable.
Simulation details are given in App.~\ref{app:details_cliff}.

The critical point $p_c$ and the correlation-length exponent $\nu$ are extracted from the scaling ansatz
\begin{equation}
    \begin{split}
        Q &= F(x) + N^{-\omega} G(x), \\
        x &= \big[ (p - p_c) + b \, (p - p_c)^2 \big] N^{1/\nu} ,
    \end{split}
    \label{eq:fss_ansatz}
\end{equation}
where $\omega$ is the leading correction-to-scaling exponent and $b$ the quadratic term of the scaling variable. For the PWR2 graph, the next correction $N^{-2\omega}H(x)$ is also included (App.~\ref{app:fss}).
We collapse the second-moment ratio of the QCMI,
\begin{equation}
    Q = \frac{\langle I\rangle^2}{\langle I^2\rangle},
\end{equation}
rather than the QCMI itself: being a ratio of moments, $Q$ is invariant under any multiplicative rescaling of the observable and therefore isolates $p_c$ and $\nu$ without reference to the order-parameter exponent $\beta$.
The collapse is performed by Gaussian-process evidence maximization~\cite{harada2011bayesian,harada2015corrections}, in which the master curve $F$ is not parameterized: $p_c$, $\nu$, $\omega$, and the kernel hyperparameters are all obtained by maximizing the log marginal likelihood.
A representative collapse is shown in Fig.~\ref{fig:phase_diagram}(a): for $\alpha=1.95$ ($d_s=40/19$), the top panel shows the raw ratio $Q$ and the bottom panel the corrected ratio $Q - N^{-\omega}G$ against $x$.
Including the leading corrections proved essential, as the accessible system sizes are not in the asymptotic limit.
The full procedure
is described in App.~\ref{app:fss}.

Figure~\ref{fig:phase_diagram}(b) collects the results, obtained from $10^4$--$10^6$ realizations per $(N,p)$ (cf.\ Table~\ref{tab:datasets}), for $128 \le N \le 2048$ (up to $4096$ at $d_s = 10/3$) and $d_s = 40/19,20/9,5/2,20/7,10/3$.
The top panel reports $p_c$ for the LR and PWR2 graphs as a function of $d_s$; the star marks the theoretical value $p_c = 0$ at the threshold spectral dimension $d_s^* = 2$.
The two families yield different values of $p_c$ as they depend on microscopic details of the model.
The critical exponents $1/\nu$ (bottom panel) are compatible
at equal $d_s$.
This agreement identifies $d_s$ as the universal quantity controlling the transition.
Moreover, the exponents of both graph families are compatible, within about $2\sigma$, with the long-range percolation values of Gori et al.~\cite{gori2017onedimensional}.
This places both models in the long-range percolation universality class.
The quoted uncertainties combine the statistical error from a joint bootstrap with the systematic error from the dependence on the fitting window and on the smallest size retained (App.~\ref{app:fss}).

Panel (c) shows two representative circuit realizations: the tableau and the two largest entanglement clusters (Sec.~\ref{sec:stab_formalism_percolation}) for $p<p_c$ and $p>p_c$.
For $p < p_c$, in the volume-law phase, the system percolates: Pauli-string weight spreads across the tableau, and a large spanning cluster coexists with smaller ones.
For $p > p_c$, in the area-law phase, the system localizes: Pauli-string weight remains concentrated, and even the largest clusters are $O(1)$.

\section{Application to quantum error correction}
\label{sec:qec}

\begin{figure*}[t]
    \centering
    \includegraphics[width=\textwidth]{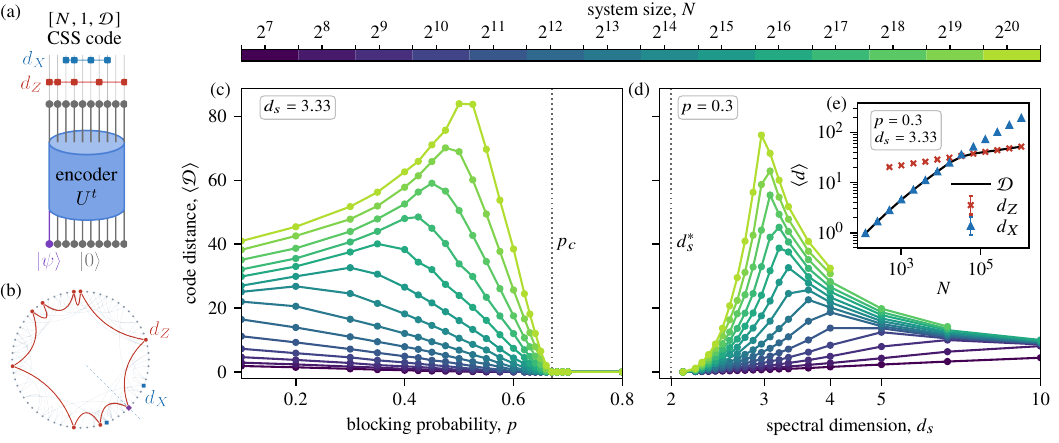}
    \caption{\textbf{CSS codes from Floquet Clifford circuits.}
    (a) The circuit as an encoder: a single data qubit $\ket{\psi}$ and $N-1$ ancillas $\ket{0}$ are evolved by $U^{t}$, defining an $[[N,1,\mathcal{D}]]$ CSS code whose distance splits into two sectors, $\mathcal{D} = \min(d_X, d_Z)$.
    (b) Representative realization of the non-blocking cluster on the ring (gray). The logical $\bar{Z}$ is a cycle winding once around the ring, of length $d_Z$ (red); the logical $\bar{X}$ is a set of sites severing every winding path, of size $d_X$ (blue).
    (c) Mean code distance $\langle \mathcal{D} \rangle$ versus blocking probability $p$ at fixed spectral dimension $d_s = 10/3$ ($\alpha = 1.6$ LR graph).
    The distance vanishes above the percolation threshold $p_c \simeq 0.67$ (dotted) and grows with $N$ below it. At finite $N$, it peaks at an optimal $p^{\rm opt} < p_c$ that approaches $p_c$ from below as the system grows.
    (d) Mean code distance $\langle \mathcal{D} \rangle$ versus spectral dimension at fixed $p = 0.3$.
    The distance vanishes below the threshold spectral dimension $d_s^*=2$ (dotted) and grows with $N$ above it. At finite $N$, it peaks at an optimal value that decreases toward $d_s^*$ with $N$.
    (e) The two sectors versus system size at $p = 0.3$ and $d_s = 10/3$: $\mathcal{D}$ is limited by the cosystole $d_X$ for $N < N^{*}$ and by the systole $d_Z$ for $N > N^{*}$, with $N^{*} \simeq 3 \times 10^{4}$ at these parameters.
    Realizations without a winding cycle carry no logical $\bar{Z}$ and are excluded from $\langle d_Z \rangle$, which is shown only for $N \ge 512$, where at least $99\%$ of realizations percolate. Averages are over $10^{3}$ realizations for $N \le 2^{14}$ and $10^{2}$ above; error bars are smaller than the symbols.
    }
    \label{fig:css_codes}
\end{figure*}
The quantum circuits introduced in Sec.~\ref{sec:model} define stabilizer codes~\cite{gottesman1996qec,calderbank1998QEC}.
Initializing one qubit in a logical state $\ket{\psi}$ and the remaining $N-1$ ancillas in $\ket{0}$, the evolution $E = U^t$ acts as the encoding map of an
$[[N,1,\mathcal{D}]]$ stabilizer code [Fig.~\ref{fig:css_codes}(a)], whose distance $\mathcal{D}$ is the minimum weight of a nontrivial logical operator.

Choosing the non-blocking gate NB $= \mathrm{CNOT}$ and the blocking gate
B $= \id$ (i.e., no gate) yields a Calderbank--Shor--Steane (CSS) code~\cite{calderbank1996CSS,steane1996CSS}.
This choice allows rigorous statements about the scaling of $\mathcal{D}$, and it affects only microscopic details, while the universal behavior is
controlled by the spectral dimension $d_s$ of the underlying graph.

In a CSS code, $X$- and $Z$-type stabilizers do not mix, so every nontrivial logical operator is purely of one type and the minimization defining $\mathcal{D}$ splits into two independent sectors,
\begin{equation}
    \mathcal{D} \;=\; \min\left( d_X, \, d_Z \right) ,
    \label{eq:d_min}
\end{equation}
with $d_X$ and $d_Z$ the minimum weights of nontrivial $X$- and $Z$-type logicals.
When the graph is embedded in a ring geometry, each sector has a geometric meaning, the two being dual to one another.
A $Z$-type operator commutes with all stabilizers only if it closes into a cycle of the ring --- a contractible cycle is itself a stabilizer.
Only cycles winding an odd number of times around the ring are nontrivial logicals, so $d_Z$ is the length of the shortest winding cycle --- the \emph{systole} of the graph.
Dually, an $X$-type logical is a set of sites whose removal severs every winding path, and $d_X$ is the size of the smallest such cut --- the \emph{cosystole}. Any winding cycle and any severing cut meet an odd number of times, which is precisely the anticommutation $\bar{X}\bar{Z} = -\bar{Z}\bar{X}$ that makes the pair a logical qubit; a representative cycle and cut are shown in Fig.~\ref{fig:css_codes}(b).
Both distances are properties of the graph geometry exactly computable in polynomial time, allowing us to obtain the ensemble statistics of $\mathcal{D}$ at sizes far beyond those accessible to tableau simulation (App.~\ref{app:qec}).

\subsection{Scaling in the two phases}
The two phases --- volume-law and area-law --- yield qualitatively different codes.
The code encodes a logical qubit only when a winding cycle exists, i.e., in the volume-law phase, and encodes nothing in the localized one.
In the latter, the non-blocking cluster fragments into finite pieces and no cycle winds the ring; therefore, no logical $\bar{Z}$ exists; dually, the ring can be severed at $O(1)$ cost, since some meridian is already free of crossing bonds.
The percolation transition is therefore also the transition at which the code ceases to encode: for $p>p_c$ no logical qubit exists, however large the system, and we label $\mathcal{D}=0$ in this phase.

In the percolating phase both logicals are supported by the spanning cluster, and the two distances grow with different laws.
The cosystole $d_X$ is set by the number of long-range bonds a meridian must cut, the same cut weight entering the cutting probability in Eq.~\eqref{eq:Pc_scaling}, yielding
\begin{equation}
    d_X(N) \;\sim\; N^{1 -2/d_s} .
    \label{eq:dX_scaling}
\end{equation}
The systole $d_Z$ is set by the number of edges in the shortest path winding the graph.
For the LR graph family it can be shown~\cite{benjamini2001diameter,coppersmith2002diameter,biskup2004chemical} that for $1 < \alpha < 2$ the systole scales with a poly-logarithmic law
\begin{equation}
    d_Z(N) \;\sim\; \left( \ln N \right)^{\Delta} ,
    \qquad
    \Delta^{-1} = \log_2\!\left( \frac{2}{\alpha} \right).
    \label{eq:dZ_scaling}
\end{equation}
It interpolates between $\Delta \to 1$ as $\alpha \to 1^{+}$ ($d_Z \sim \ln N$) and $\Delta \to \infty$ as $\alpha \to 2^{-}$, where
shortcuts become rare, and the distance crosses over to
polynomial growth.

Varying the geometry reveals a trade-off between $d_X$ and $d_Z$.
In a fully connected graph, separating the system into two disconnected parts requires removing an extensive number of vertices, yielding $d_X = O(N)$.
At the same time, winding around the graph requires only $d_Z\sim \ln N$ steps.
This logarithmic scaling is a hallmark of the small-world phenomenon~\cite{watts1999smallworld}, and is associated with an effectively infinite spectral dimension in the corresponding class of small-world networks.
In contrast, a nearest-neighbor ring can be disconnected by removing a single site, giving $d_X = O(1)$, while a nontrivial winding path necessarily spans the entire system, yielding $d_Z = O(N)$.

The spectral dimension provides a general framework, organizing graph geometries along a continuum between these two extremes and highlighting the trade-offs underlying the code distance $\mathcal{D}$.
The LR and PWR2 graph families provide examples of sparse architectures with only $O(1)$ average connectivity per site and a tunable and potentially large spectral dimension.
As a result, they simultaneously support a diverging cut distance $d_X$ and a diverging winding distance $d_Z$, leading to nontrivial code-distance scaling despite their sparse connectivity. This suggests that the performance of dynamically generated codes is controlled not by the average degree or graph density, but by the spectral dimension, which quantifies how efficiently connectivity is organized across scales.

\subsection{Numerical results}
We numerically study the scaling of the code distance and its two sectors by sampling LR graphs directly. For each realization, we compute the systole $d_Z$ and cosystole $d_X$ as shortest-cycle and minimum-cut problems on the graph. We first fix $d_s = 10/3$ ($\alpha=1.6$) and average over realizations to obtain $\langle \mathcal{D} \rangle$ on a grid of $p \in [0,1]$ and system sizes $N = 2^{7}$ to $2^{20}$ [Fig.~\ref{fig:css_codes}(c)]. We then fix $p=0.3$ and repeat the analysis on a grid $d_s \in [2,10]$ [Fig.~\ref{fig:css_codes}(d)].

The data show two distinct regimes. At fixed $d_s=10/3$, $\mathcal{D}$ vanishes in the thermodynamic limit for $p>p_c$, while it grows with $N$ for $p<p_c$ [Fig.~\ref{fig:css_codes}(c)]. Similarly, at fixed $p=0.3$, the distance vanishes below a threshold spectral dimension $d_s^*=2$ and grows with $N$ above it [Fig.~\ref{fig:css_codes}(d)].

The non-monotonic behavior of the distance at finite $N$ results from the competition between its two sectors. Raising $p$ (or lowering $d_s$) removes long shortcuts, making winding paths more costly and hence increasing $d_Z$. At the same time, it thins the bonds crossing a meridian, making cuts cheaper and decreasing $d_X$. Thus, $\mathcal{D}=\min(d_X,d_Z)$ is largest near the crossing of the two sectors. At fixed $N$, this produces an optimal $p^{\rm opt}(N)<p_c$ in panel (c), which approaches $p_c$ from below as $N$ increases. Likewise, in panel (d), the finite-size maximum occurs at an optimal $d_s^{\rm opt}(N)>d_s^*$, which decreases toward $d_s^*$ with increasing $N$.

The competition between the two sectors is apparent as a function of system size [Fig.~\ref{fig:css_codes}(e)]. At $p=0.3$ and $d_s=10/3$, the distance is limited by the cosystole $d_X$ for $N<N^*$ and by the systole $d_Z$ for $N>N^*$, with $N^*\simeq 3\times10^4$. This inversion occurs because $d_X$ grows as a power law with $N$, whereas $d_Z$ grows only poly-logarithmically, so that the systole eventually becomes the limiting sector.
The inversion size $N^*=N^*(d_s,p)$ depends on both parameters, with distinct geometric origins. At fixed $d_s$, increasing $p$ dilutes the non-blocking cluster: shortcuts become scarce, $d_Z$ increases, and $N^*$ grows monotonically with $p$. At small $p$, the connectivity is dense, and the systole sets the minimum already at modest sizes.
At fixed $p$, varying $d_s$ instead changes the edge-length distribution. At large $d_s$, long shortcuts are abundant, $d_Z$ is small, and the systole again dominates at modest sizes. As $d_s\to 2$, the connectivity becomes effectively short-ranged, severing the ring becomes cheap, $N^*$ diverges, and $\mathcal{D}\simeq d_X$ over the accessible range.

The distance of these codes is thus a geometric quantity: it is induced by the connectivity of the graph and undergoes a transition at the percolation threshold $p_c$.
Long-range connectivity is a tunable resource in several physical platforms, and our results show that its optimal value depends on the system size. Geometry enters the code twice, with opposite effects: the systole is set by the availability of shortcuts, while the cosystole is set by the cost of severing the ring. The distance therefore peaks at an optimal $p^{\rm opt}(N)$ rather than increasing monotonically.

For a given platform, the accessible range of connectivity and number of qubits define a region of the $(d_s, N)$ plane. The remaining parameter $p$ can then be tuned to maximize the code distance within this region. The CSS choice NB $= \mathrm{CNOT}$, B $= \id$ makes this analysis particularly transparent: it splits the distance into two sectors with independent geometric meanings, each computable directly on the sampled graph. For a general pair B, NB, the sectors no longer decouple and these geometric quantities become bounds, but the same trade-off is expected to persist: shortcuts and cuts respond oppositely to dilution, leading to an optimal connectivity.

\section{Discussion \& Outlook}
This work establishes the spectral dimension as the fundamental geometric quantity governing quantum information dynamics on complex interaction graphs. While the underlying graph geometries may differ drastically in terms of connectivity, diameter, or embedding dimension, their entanglement dynamics become universal once characterized by the same spectral dimension.
Therefore, in graph-based quantum dynamics, $d_s$ plays the same role of spatial dimension in conventional critical phenomena.

Our analysis reveals a universal geometric criterion for the existence of an entanglement transition on random graphs.
Relating localization to the probability of disconnecting the graph through blocking cuts, we identified the threshold spectral dimension $d_s^*=2$ separating graph geometries that are always localized from those capable of supporting a transition between area-law and volume-law entangled phases.
Above this threshold, the transition is governed by the percolation of ergodic clusters and inherits the universality class associated with the spectral properties of the graph.

This picture implies a universality that depends only on the low-energy Laplacian spectrum and not on microscopic connectivity.
As a result, graph families with exponentially different connectivities can realize the same entanglement transition. For example, the power-of-two graph reproduces the universality class of long-range random graphs while requiring only logarithmically many connections per site. More generally, our results suggest that spectral engineering may provide a systematic route to designing desired quantum-information dynamics using sparse and experimentally realizable architectures.

The implications of our results extend beyond entanglement dynamics and provide a complementary perspective on quantum error correction.
In our construction, graph geometry determines both the existence and the quality of the dynamically generated CSS code.
Below the percolation threshold, the circuit encodes a logical qubit with code distance growing with system size, whereas above $p_c$ the logical subspace disappears altogether. The connection to quantum error correction provides a complementary interpretation of the role of graph geometry. We showed that the spectral dimension controls the competition between the two logical sectors entering the code distance. Increasing connectivity enhances the cost of disconnecting the graph and therefore improves protection against one class of logical errors, while simultaneously introducing shortcuts that reduce the length of nontrivial logical cycles. The spectral dimension organizes this trade-off by interpolating continuously between local Euclidean lattices and highly connected small-world-type graphs.
From this perspective, interaction graphs with intermediate spectral dimension appear as natural candidates for optimizing code performance under locality constraints. Rather than focusing on connectivity alone, our results suggest that spectral dimension may provide a useful design principle for dynamically generated quantum codes and more general graph-based quantum-information architectures. In this sense, $d_s$ offers a design principle complementary to existing expander-based constructions, allowing one to systematically explore sparse graph geometries that balance physical locality with favorable code distance scaling.

Different directions open as outlook for future research. First, it would be interesting to understand how the present framework extends beyond CSS codes and whether spectral dimension similarly constrains more general QEC codes, including qLDPC, subsystem, non-CSS, and finite-rate constructions.
Moreover, introducing a finite density of non-Clifford gates would provide a natural route toward genuinely quantum chaotic dynamics. In analogy with previous studies of Floquet localization, one may expect the sharp blocking structure of the Clifford limit to broaden into a finite localization length while preserving the underlying geometric mechanism \cite{chandran2015semiclassical}.

\section*{Acknowledgments}
We acknowledge fruitful discussions with Matthew P.\ A.\ Fisher.
The authors used Claude Opus 4.8/5/5.5 during the preparation of this manuscript to improve readability and assist with code optimization.
The authors assume full responsibility for the scientific content, accuracy, and integrity of this paper.
Funded by the Deutsche Forschungsgemeinschaft (DFG, German Research
Foundation) -- 544919793, and the European Union (ERC, QuSimCtrl, 101113633).
Views and opinions expressed are, however, those of the authors only and do not necessarily reflect those of the European Union or the European Research Council Executive Agency. Neither the European Union nor the granting authority can be held responsible for them. A.S. acknowledges support from the Alexander von Humboldt Foundation through a Humboldt research fellowship for postdoctoral researchers.

\textit{Data availability statement.---}%
The data and code associated with this manuscript are deposited in Zenodo \href{https://doi.org/10.5281/zenodo.23184962}{10.5281/zenodo.23184962}.
Automatic access is restricted; access credentials will be provided to human researchers upon reasonable request via Zenodo for academic and verification purposes.

\clearpage
\begin{appendix}
\resumetoc
\tableofcontents

\section{Blocking and non-blocking gates}
\label{app:BNB_gates}
The Clifford circuits considered in this work are composed of blocking (B) and non-blocking (NB) gates: two-qubit Clifford gates $U \in \mathcal{C}_2$ that either block or transmit quantum information, respectively.
The defining feature of a blocking gate $U_{\rm B}$ is that it possesses a local integral of motion (LIOM), i.e., a \emph{single-site} operator $O$ satisfying $[U_{\rm B}, O]=0$, which prevents information from spreading across the edge.
In contrast, an NB gate $U_\mathrm{NB}$ scrambles local operators, allowing information and entanglement to spread.
While we use two specific gate realizations, our results depend only on the abstract B/NB structure and are independent of their microscopic implementation.
Here, we characterize the B and NB gates, formalize in which sense B gates are ``blocking,'' and show the generality of this construction.

The B and NB gates were defined by their action on the Pauli group: any two-qubit Clifford gate is fixed, up to a global phase, by the images of $ZI$, $XI$, $IZ$, $IX$.
We retrieve the associated unitary transformation requiring $U P U^{\dagger} = P'$ for each Pauli string $P$.
For the pair of Sec.~\ref{sec:model} one finds:
\begin{equation}
\begin{split}
    &U_{\rm B} = \mathrm{CZ} = \mathrm{diag}(1,1,1,-1), \\
    &U_{\rm NB} = \frac{1}{\sqrt{2}}
    \begin{pmatrix}
        i & 1 & 0 & 0\\
        0 & 0 & -1 & i\\
        0 & 0 & -1 & -i\\
        i & -1 & 0 & 0
    \end{pmatrix}.
\end{split}
\label{eq:gate_unitaries}
\end{equation}
Both unitaries are determined only up to sixteen sign choices, but none of the statements below depends on the choice.

\begin{figure}[t]
    \centering
    \includegraphics{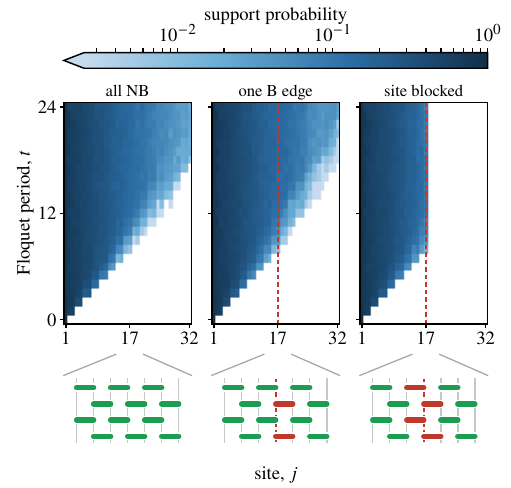}
    \caption{\textbf{Blockade is a property of a site, not of an edge.}
    Probability that the operator $Z_1(t)$, seeded at the leftmost edge of an $N=32$ 1D chain, acts nontrivially on site $j$ after $t$ periods of the brick-wall Floquet circuit.
    Each realization draws an independent random two-qubit Clifford with no single-site LIOM for every NB gate, while the B gate is fixed to CZ; the average is over $400$ realizations.
    The color scale is logarithmic and starts at one realization; sites outside the support in every realization are left white.
    Below each plot, schematic zooms on seven sites around the dashed one show two Floquet periods, with NB gates (green) and B gates (red).
    \emph{Left:} with only NB gates, the operator support spreads ballistically.
    \emph{Middle:} a single B gate at the dashed line delays the front but does not stop it.
    \emph{Right:} when both edges adjacent to the dashed site are blocking, the front stops there for all times.
    }
    \label{fig:gates}
\end{figure}
A single edge with a B gate does not by itself block information propagation.
Although the B gate has a conserved operator supported on its two endpoints, another edge connected to either endpoint can transport this operator away.
Blockade is therefore a property of a site: all edges incident on the site must be blocking, such that a common local operator $O$ commutes with every gate acting on those edges and is strictly conserved.
To illustrate this, we construct a brick-wall circuit on an $N=32$ 1D chain with open boundary conditions.
We consider three scenarios: all edges contain an NB gate, only one edge contains a B gate, and both edges touching a site contain a B gate.
We schematically represent two Floquet periods below the plots in Fig.~\ref{fig:gates} with NB gates in green and B gates in red.
We sample different realizations by drawing an independent random two-qubit Clifford with no single-site LIOM for each NB gate, while the B gate is fixed to CZ as in Eq.~\eqref{eq:gate_unitaries}.
We then compute the probability that the operator $Z_1(t)$, seeded at the leftmost edge, acts nontrivially on site $j$ after $t$ Floquet periods.
The results are shown in Fig.~\ref{fig:gates}.
Operator spreading occurs both when all gates are NB and when only one bond carries a B gate --- the latter partially slows down the propagation.
In contrast, a site where exclusively B gates are applied completely blocks information spreading.
This demonstrates why the criterion of Sec.~\ref{sec:entanglement_t_and_cut_prob} is formulated per site.

Note that B gates are not trivially single-qubit basis rotations.
Every two-qubit Clifford can be written as $L_1 V L_2$ with $L_i$ local and $V \in \{\id, \mathrm{CNOT}, \mathrm{iSWAP}, \mathrm{SWAP}\}$~\cite{kraus2001twoqubit,khaneja2001cartan}, and searching over all $576 \times 576$ pairs of local Cliffords places $U_{\rm B}$ in the CNOT class and $U_{\rm NB}$ in the iSWAP class.
This assignment is, however, incidental: the local-equivalence class does not decide whether a gate blocks.
The CNOT class contains $U_{\rm B} = \mathrm{CZ}$ and the CNOT itself, and the CSS implementation of Sec.~\ref{sec:model} uses the latter as its NB gate.

Both B and NB gates are entangling.
In particular, $U_{\rm B}$ takes $|{+}{+}\rangle$ to a maximally entangled state, and of the $36$ two-qubit stabilizer product states~\footnote{The single-qubit stabilizer states are the six Pauli eigenstates $|0\rangle, |1\rangle, |{\pm}\rangle, |{\pm}i\rangle$; their $6\times 6 = 36$ products are the product states among the $60$ two-qubit stabilizer states, the remaining $24$ being maximally entangled.
For two qubits, a stabilizer state is either product or maximally entangled, so this count characterizes the gate's action on them completely.}
each of $U_{\rm B}$ and $U_{\rm NB}$ entangles exactly $16$.
Entangling power is therefore not the axis that separates the two.

Modulo global phase, a Clifford unitary is fully specified by the images of $\{ X_i, Z_i \}$, each being a Pauli string with a sign.
For $U \in \mathcal{C}_2$ there are $4$ such images and hence $2^4=16$ sign assignments.
Dressing a Clifford by Pauli operators changes signs but not where an operator is supported:
for $U' = PU$ with $P \in \mathcal{P}_2$ a Pauli, $U' Q U'^\dagger = P (UQU^\dagger) P^\dagger = \pm U Q U^\dagger$ because two Paulis commute or anticommute.
Therefore, the blocking property is a property of the $11520 / 16 = 720$ symplectic classes, and we can characterize the whole group.
In doing so we also need to take into account the orientation of the two-qubit gate.
In fact, a gate may leave a single-site operator invariant at one end, at the other, at both, or at neither.
The whole group is partitioned into six families summarized in Table~\ref{tab:gate_families}.
For each gate, we tested its behavior numerically as above --- seed every Pauli generator to the left of $m$, accumulate the span, and ask whether it ever reaches the right of $m$.

\begin{table}[b]
    \caption{The $720$ symplectic classes, sorted by the single-site operators
    left invariant at each end of the gate, by the orientations for which a
    site $m$ is blocked, and by whether the blockade survives a site of
    degree larger than two.}
    \label{tab:gate_families}
    \begin{ruledtabular}
    \begin{tabular}{clccc}
        & family & \# & blocks when & degree \\
        \hline
        \multirow{2}{*}{\rotatebox[origin=c]{90}{trivial}}
        & locally equiv.\ to $\id$  & $36$  & always & any
        \rule{0pt}{1.3em} \\
        & locally equiv.\ to SWAP   & $36$  & never  & ---
        \rule[-1.1em]{0pt}{0pt} \\
        \hline
        \multirow{6}{*}{\rotatebox[origin=c]{90}{entangling}}
          & \multicolumn{4}{l}{\emph{invariant single-site operator at}} \\
          & both ends, the same one   & $12$  & any orientation     & any \\
          & both ends, different ones & $24$  & all edges same end  & any \\
          & one end only              & $144$ & all edges that end  & any \\
          & neither end               & $24$  & the ends alternate  & even \\
          & neither end               & $444$ & never               & --- \\
    \end{tabular}
    \end{ruledtabular}
\end{table}

\textit{Locally equivalent to the identity ($36$).} ---
These are the product gates $L_1 \otimes L_2$.
They do not couple the two sites at all, so nothing crosses $m$ whatever the orientation and whatever the degree.
They block for a trivial reason and carry no physics.

\textit{Locally equivalent to SWAP ($36$).} ---
These are the remaining non-entangling classes, and they never block: SWAP carries every operator from one site to the other and fixes nothing at either end.
Together with the previous family, they exhaust the $72$ non-entangling classes, leaving $648$ genuinely entangling ones.

\textit{The same invariant operator at both ends ($12$).} ---
This is the definition adopted in the main text, and CZ belongs here.
They form three families of four, one for each choice of conserved single-site Pauli $X$, $Y$ or $Z$.
Because the operator is fixed at either end, the blockade is insensitive to how the gates are laid on the edges and to how many edges meet at $m$.

\textit{Different invariant operators at the two ends ($24$).} ---
Six families of four, one for each ordered pair of distinct Paulis.
Such a gate blocks only when every edge at $m$ presents the same end, so that one operator is fixed by all of them.
The CNOT is the familiar member, fixing $Z$ on its control and $X$ on its target: a site that is the control of all its edges --- or the target of all of them --- is blocked, while the mixed, and generic, case is not.
This is precisely why the randomly oriented CNOT of the CSS implementation of Sec.~\ref{sec:model} is non-blocking.
Once the orientation is chosen consistently, the blockade is again robust to the degree.

\textit{An invariant operator at one end only ($144$).} ---
Twenty-four classes for each of the six choices of end and operator.
Only that end can produce a blockade, and it does so under the same condition as above: every edge at $m$ must present it.
The blockade is once more degree-robust.

\textit{No invariant operator at either end ($24 + 444$).} ---
Of these, $444$ never block.
The remaining $24$ do, and are the only classes that block without any gate-level invariant. The class
\begin{equation}
    XI \mapsto ZZ, \quad IX \mapsto IZ, \quad ZI \mapsto XI, \quad IZ \mapsto XX
    \label{eq:noliom_blocker}
\end{equation}
fixes no single-site Pauli at either end, yet each end transmits exactly one, and the two ends transmit different ones: $Z_m \mapsto X_m$ at one end and $X_m \mapsto Z_m$ at the other.
Equivalently, these gates conserve a single-site operator at $m$ over a complete Floquet period, but not gate by gate.
The alternation of the ends must then close on itself along the order in which the circuit visits the edges of $m$: at degree $z$ this is impossible for odd $z$, while for even $z$ it fixes the orientation up to a global flip, leaving exactly two of the $2^z$ assignments.
At degree two the brick-wall circuit supplies one of them for free, which is what makes these gates look like ordinary blockers on the chain.

Since the sites of the long-range graphs studied here carry arbitrary degree and the gates are placed independently on each edge, the criterion adopted in Sec.~\ref{sec:model} is not merely a convenient sufficient condition: it is the only one that survives.

\section{Cutting probability: derivation and scaling}
\label{app:cutting_p}

Consider $N$ qubits labeled $1,\dots,N$ embedded in a one-dimensional Euclidean space, with interactions forming a general graph in which each pair $(a,b)$ is connected with probability $p_{a,b}=p_{b,a}\in[0,1]$; we set $p_{a,a}=0$.
Every realized edge hosts one two-qubit gate, drawn independently as blocking (B) with probability $p$ and non-blocking (NB) with probability $1-p$ (cf.\ Sec.~\ref{sec:model}).
Site $i$ forms an \emph{absolute cut} when every realized edge that straddles or touches it --- every edge $(a,b)$ with $a<b$ and $a \le i \le b$ --- carries a blocking gate.
The non-blocking subgraph then leaves $i$ isolated and disconnects $\{a<i\}$ from $\{b>i\}$.
Everything that follows holds for any symmetric $p_{a,b}$: the embedding enters only through the labels, which decide which edges straddle or touch a given site.
Additional assumptions are introduced, and marked as such, where they are used.

We first compute the probability $P_{\rm cut}(i)$ that a given site $i$ hosts an absolute cut.
Since edges and gates are independent, an edge $(a,b)$ hosts an NB gate with probability $\overline{p}_{a,b} := p_{a,b}(1-p)$ and it either carries a blocking gate or is absent with probability $1-\overline{p}_{a,b}$.
We assume $\overline{p}_{\max} := \max_{a,b}\overline{p}_{a,b} < 1$.
If $\overline{p}_{a,b}=1$ then $p=0$ and $p_{a,b}=1$, and $P_{\rm cut}(i)=0$ for every $a\le i\le b$.
Only the edges in
\begin{equation}
    E_i := \{(a,b) : a<b,\; a\le i\le b\}
    \label{eq:Ei}
\end{equation}
contribute to $P_{\rm cut}(i)$.
Writing their endpoints as $a=i-\ell$ and $b=i+j$, we obtain
\begin{equation}
    P_{\rm cut}(i) = \prod_{(a,b)\in E_i}\left(1-\overline{p}_{a,b}\right)
    = \prod_{\ell =0}^{i-1} \ \prod_{j=0}^{N-i}
        \left( 1 - \overline{p}_{i-\ell, \, i+j} \right),
    \label{eq:Pcut-product-appendix}
\end{equation}
the term $\ell=j=0$ contributing a factor one since $\overline{p}_{i,i}=0$.
In additive form,
\begin{equation}
   -\ln P_{\rm cut}(i)
   = -\sum_{(a,b)\in E_i} \ln\!\left( 1 - \overline{p}_{a,b} \right).
   \label{eq:ln_Pb}
\end{equation}

Wanting to rewrite $P_{\rm cut}$ in terms of a bipartition, we note that every edge of $E_i$ has either $b>i$, or $b=i$.
Hence $E_i$ is the disjoint union
\begin{equation}
\begin{split}
    &E_i = E(S_i)\,\sqcup\,F_i, \\
    &E(S_i) := \{(a,b): a\le i<b\}, \\
    &F_i := \{(a,i): a<i\}.
\end{split}
\label{eq:Ei-split}
\end{equation}
The edges in $E(S_i)$ are those severed by the bipartition of the chain into $S_i=\{1,\dots,i\}$ and its complement.
The edges in $F_i$ join $i$ to sites on its left and lie inside $S_i$.
We denote the total weights of the two sets by
\begin{equation}
    \mathrm{Cut}(S_i) := \sum_{a\le i<b}\overline{p}_{a,b},
    \qquad
    d_i^- := \sum_{a<i}\overline{p}_{a,i},
    \label{eq:cut-def}
\end{equation}
so that $\sum_{(a,b)\in E_i}\overline{p}_{a,b}=\mathrm{Cut}(S_i)+d_i^-$.
$\mathrm{Cut}(S_i)$ is the expected number of NB gates on edges crossing the bipartition.
Since every edge incident to $i$ belongs to $E_i$ --- the edges $(a,i)$ with $a<i$ to $F_i$, the edges $(i,b)$ with $b>i$ to $E(S_i)$ ---, also
$\mathrm{Cut}(S_i)+d_i^-\ge\mathbb{D}_{ii}$, with $\mathbb{D}_{ii}$ the annealed degree of $i$ defined below.

We can now express $P_{\rm cut}(i)$ in terms of the spectral properties of the interaction graph.
Let $\chi_i \in \{0,1\}^N$ denote the indicator of $S_i$,
\begin{equation}
    (\chi_i)_a =
    \begin{cases}
        1, & a \le i, \\
        0, & a > i,
    \end{cases}
    \label{eq:indicator}
\end{equation}
so that an edge $(a,b)$ with $a<b$ belongs to $E(S_i)$ exactly when $(\chi_i)_a \neq (\chi_i)_b$.
The relevant spectral object is the Laplacian of the annealed graph [cf.\ Eq.~\eqref{eq:laplacian-def}],
\begin{equation}
    \mathbb{L} = \mathbb{D} - \mathbb{A}, \qquad
    \mathbb{A}_{a,b} = \overline{p}_{a,b}, \qquad
    \mathbb{D}_{aa} = \sum_{b} \overline{p}_{a,b}.
    \label{eq:laplacian-app}
\end{equation}
$\mathbb{L}$ is real and symmetric, so it has a real orthonormal basis of eigenvectors.
Each row of $\mathbb{L}$ sums to zero, $\mathbb{L}\mathbf{1}=0$, and by symmetry also $\mathbf{1}^T\mathbb{L}=0$.
For any vector $y$, using $\mathbb{D}_{aa}=\sum_b\mathbb{A}_{a,b}$, the symmetry of $\mathbb{A}$, and $\mathbb{A}_{a,a}=0$,
\begin{equation}
\begin{aligned}
    y^{T} \mathbb{L}\, y
    &= \sum_{a,b}\mathbb{A}_{a,b}\,y_a^2-\sum_{a,b}\mathbb{A}_{a,b}\,y_ay_b \\
    &= \frac12\sum_{a,b}\mathbb{A}_{a,b}\left(y_a^2+y_b^2-2y_ay_b\right)\\
    &= \frac12 \sum_{a,b} \mathbb{A}_{a,b} \,(y_a - y_b)^2 \\
    &= \sum_{a<b} \overline{p}_{a,b} \,(y_a - y_b)^2 .
\end{aligned}
\label{eq:laplacian-quadratic}
\end{equation}
In particular $\mathbb{L}$ is positive semidefinite, so its eigenvalues are non-negative, and $0$ is an eigenvalue with eigenvector $\mathbf{1}$.
We order the eigenpairs $\mathbb{L}\,u_n = \lambda_n u_n$ as
$0 = \lambda_0 < \lambda_1 \le \lambda_2 \le \dots \le \lambda_{N-1} =: \lambda_{\max}$, with $u_0 = \mathbf{1}/\sqrt{N}$.
If an eigenvalue is degenerate, the eigenbasis within its eigenspace is not unique.
Every quantity used below is either an eigenvalue or a sum over complete eigenspaces, and does not depend on this choice.

Setting $y=\chi_i$ in Eq.~\eqref{eq:laplacian-quadratic}, $(y_a-y_b)^2$ equals one if $(a,b)\in E(S_i)$ and zero otherwise, so the cut weight is a Laplacian quadratic form,
\begin{equation}
    \mathrm{Cut}(S_i) = \chi_i^{T} \mathbb{L}\, \chi_i .
    \label{eq:cut-quadratic-form}
\end{equation}
The uniform mode $u_0$ carries no cut weight.
Projecting it out,
\begin{equation}
    w_i := \chi_i - \frac{i}{N}\,\mathbf{1},
    \label{eq:wi}
\end{equation}
leaves the quadratic form unchanged, $\chi_i^T\mathbb{L}\chi_i = w_i^T\mathbb{L}w_i$, because $\mathbb{L}\mathbf{1}=0$ and $\mathbf{1}^T\mathbb{L}=0$.
The vector $w_i$ is orthogonal to $\mathbf{1}$, since $\langle \mathbf{1},w_i\rangle = i-\frac{i}{N}N=0$, and its norm is
\begin{equation}
    \|w_i\|^2 = i\left(1-\frac{i}{N}\right)^2+(N-i)\left(\frac{i}{N}\right)^2
    = x(1-x)\,N ,
    \label{eq:wnorm}
\end{equation}
with $x:=\frac{i}{N}$.
Expanding in the eigenbasis, $w_i=\sum_n c_n u_n$ with $c_n := \langle u_n, w_i\rangle$.
Orthogonality to $\mathbf{1}=\sqrt{N}\,u_0$ gives $c_0=0$, and $c_n=\langle u_n,\chi_i\rangle$ for $n\ge1$.
By orthonormality,
\begin{equation}
    \sum_{n\ge1} c_n^2 = \|w_i\|^2,
    \qquad
    \mathrm{Cut}(S_i) = w_i^T\mathbb{L}\,w_i = \sum_{n \ge 1} \lambda_n \, c_n^2 .
    \label{eq:cut-spectral}
\end{equation}
Each coefficient $c_n^2$ is the weight that the step $\chi_i$ places on mode $n$.
The cut weight is the sum of these weights, each multiplied by the energy of its mode.

\subsection{Upper bound}
\label{app:upper}
Since $-\ln(1-y) \ge y$ for $y \in [0,1)$, Eqs.~\eqref{eq:ln_Pb} and~\eqref{eq:Ei-split} give
\begin{equation}
    -\ln P_{\rm cut}(i) \;\ge\; \mathrm{Cut}(S_i)+d_i^- \;\ge\; \mathrm{Cut}(S_i) = \chi_i^{T} \mathbb{L}\, \chi_i .
    \label{eq:P-cut-bound}
\end{equation}
The edges in $F_i$ do not cross $S_i$ and do not contribute to $\chi_i^T\mathbb{L}\chi_i$.

Since $c_0=0$, every eigenvalue left in the sum of Eq.~\eqref{eq:cut-spectral} is $\ge \lambda_1$, and
\begin{equation}
    \mathrm{Cut}(S_i) = \sum_{n \ge 1} \lambda_n \, c_n^2
    \;\ge\; \lambda_1 \sum_{n\ge1} c_n^2 = \lambda_1 \|w_i\|^2 .
    \label{eq:fiedler}
\end{equation}
Combining with Eqs.~\eqref{eq:P-cut-bound} and~\eqref{eq:wnorm},
\begin{equation}
    P_{\rm cut}(i) \;\le\; \exp\!\left( - x(1-x)\, N \lambda_1 \right).
    \label{eq:Pb-upper-bound-gap}
\end{equation}
This bound is general and holds for any graph.
It shows that absolute cuts at any fixed fraction $x\in(0,1)$ of the chain disappear whenever $N\lambda_1\to\infty$.

To express it through the spectral dimension, we use Eq.~\eqref{eq:ds-gap} as a lower bound on the gap: there is a constant $g>0$, independent of $N$, such that $\lambda_1 \ge g\,N^{-2/d_s}$.
Then
\begin{equation}
    P_{\rm cut}(i) \;\le\; \exp\!\left( - g\,x(1-x)\, N^{\,1 - 2/d_s} \right).
    \label{eq:Pb-upper-bound}
\end{equation}
For $d_s > 2$, absolute cuts disappear exponentially in system size.
Localization can then no longer be sustained by isolated cuts and, if present, must emerge collectively, from extended clusters of blocking bonds.
For $d_s \le 2$, $N\lambda_1 = O(1)$ and Eq.~\eqref{eq:Pb-upper-bound-gap} carries no information.

\subsection{Lower bound}
\label{app:lower}
The upper bound is unconditional but one-sided: it shows that absolute cuts disappear once $d_s > 2$, and says nothing when $d_s \le 2$.
We now bound $P_{\rm cut}$ from below, which means bounding $-\ln P_{\rm cut}$, and hence $\mathrm{Cut}(S_i)$, from above.
Unlike the upper bound, this requires assumptions.

As for the upper bound, we first reduce the problem to the cut weight.
For $\overline{p}\in[0,\overline{p}_{\max}]$,
\begin{equation*}
    -\ln(1-\overline{p})=\sum_{n\ge1}\frac{\overline{p}^n}{n}\;\le\;\sum_{n\ge1}\overline{p}^n=\frac{\overline{p}}{1-\overline{p}}\;\le\;\frac{\overline{p}}{1-\overline{p}_{\max}} .
\end{equation*}
Applying this edge by edge in Eq.~\eqref{eq:ln_Pb}, and combining with Eq.~\eqref{eq:P-cut-bound},
\begin{equation}
    \mathrm{Cut}(S_i) \;\le\; -\ln P_{\rm cut}(i)
    \;\le\; \frac{\mathrm{Cut}(S_i)+d_i^-}{1 - \overline{p}_{\max}} .
    \label{eq:sandwich}
\end{equation}
The additional weight is bounded by the degree, $d_i^-\le\mathbb{D}_{ii}$, and the degree by the spectrum.
With $e_i$ the $i$-th unit vector, $\mathbb{D}_{ii}=e_i^T\mathbb{L}\,e_i\le\lambda_{\max}\|e_i\|^2=\lambda_{\max}$.
Hence
\begin{equation}
    -\ln P_{\rm cut}(i) \;\le\; \frac{\mathrm{Cut}(S_i)+\lambda_{\max}}{1 - \overline{p}_{\max}} .
    \label{eq:sandwich-2}
\end{equation}
and the problem reduces to bounding $\mathrm{Cut}(S_i)$ and $\lambda_{\max}$ from above.

For $\lambda\in\mathbb{R}$, let $V_{>\lambda}:=\mathrm{span}\{u_n:\lambda_n>\lambda\}$ be the span of the modes with energy above $\lambda$.
Its orthogonal complement $V_{\le\lambda}$ is spanned by the modes with $\lambda_n\le\lambda$.
The orthogonal projector onto $V_{>\lambda}$ is
\begin{equation}
    \Pi_{>\lambda} := \sum_{n:\,\lambda_n>\lambda} u_n u_n^T ,
\end{equation}
and $\Pi_{\le\lambda}:=\mathbb{I}-\Pi_{>\lambda}$, with $\mathbb{I}$ the identity, projects onto $V_{\le\lambda}$.
$V_{>\lambda}$ is the direct sum of the eigenspaces with eigenvalue larger than $\lambda$, so $\Pi_{>\lambda}$ depends only on $\mathbb{L}$ and not on the choice of eigenbasis.
We define the \emph{spectral tail} of the cut,
\begin{equation}
    T_i(\lambda) := \big\|\Pi_{>\lambda}\, w_i\big\|^2 = \sum_{n:\,\lambda_n>\lambda} c_n^2 .
    \label{eq:T-def}
\end{equation}
The second expression holds because $\Pi_{>\lambda}w_i=\sum_{n:\,\lambda_n>\lambda}c_n u_n$ is a combination of orthonormal vectors, whose squared norm is the sum of the squared coefficients.
$T_i(\lambda)$ is the part of the weight $\|w_i\|^2$ carried by modes of energy above $\lambda$.
It has three elementary properties:
(i) it is non-increasing in $\lambda$, since raising $\lambda$ removes terms from the sum;
(ii) $0\le T_i(\lambda)\le\|w_i\|^2\le N/4$;
(iii) for $0\le\lambda<\lambda_1$ the sum runs over all $n\ge1$, so $T_i(\lambda)=\|w_i\|^2$ by $c_0=0$ and Eq.~\eqref{eq:cut-spectral}, while for $\lambda\ge\lambda_{\max}$ the sum is empty and $T_i(\lambda)=0$.

$T_i(\lambda)$ also has a geometric meaning.
It is the squared error of the best approximation of the step $\chi_i$ by a combination of modes of energy at most $\lambda$,
\begin{equation}
    T_i(\lambda) = \min_{v\in V_{\le\lambda}} \|\chi_i - v\|^2
    \qquad (\lambda\ge0).
    \label{eq:T-distance}
\end{equation}

The cut weight is the area under the spectral tail,
\begin{equation}
    \mathrm{Cut}(S_i) = \int_0^{\lambda_{\max}} T_i(\lambda)\, d\lambda .
    \label{eq:layer-cake}
\end{equation}
In fact, since $0\le\lambda_n\le\lambda_{\max}$ for all $n$,
\begin{equation*}
    \mathrm{Cut}(S_i)=\sum_{n\ge1}c_n^2\,\lambda_n
    =\sum_{n\ge1}c_n^2\int_0^{\lambda_n}d\lambda
    =\int_0^{\lambda_{\max}}d\lambda\sum_{n:\,\lambda_n>\lambda}c_n^2 .
\end{equation*}
In the last step the finite sum and the integral have been exchanged.
A point $\lambda\in[0,\lambda_{\max}]$ lies in the integration range $[0,\lambda_n)$ of the $n$-th term exactly when $\lambda_n>\lambda$; endpoints have zero measure.
Equation~\eqref{eq:layer-cake} is the familiar identity $\mathbb{E}[X]=\int_0^\infty \Pr(X>t)\,dt$ for a non-negative random variable $X$.
After normalization, $c_n^2/\|w_i\|^2$ is a probability distribution over the eigenvalues, $\mathrm{Cut}(S_i)/\|w_i\|^2$ is its mean, and $T_i(\lambda)/\|w_i\|^2$ is its survival function.

In the following we introduce three assumptions.
Consider the integrated density of states, normalized to the fraction of modes,
\begin{equation}
    \mathcal{N}(\lambda) := \frac{1}{N}\,\#\{n : \lambda_n\le\lambda\} .
    \label{eq:IDOS}
\end{equation}
It counts the zero mode, so $\mathcal{N}(\lambda)=1/N$ for $0\le\lambda<\lambda_1$.
It is non-decreasing and equals one for $\lambda\ge\lambda_{\max}$.
In a one-dimensional embedding, $1/\mathcal{N}(\lambda)$ is the number of sites per mode of energy at most $\lambda$.
All constants below ($c$, $\lambda^\ast$, $C$, $\Lambda$) are positive and independent of $N$.

\medskip
\textbf{(A1) Soft modes.}
\emph{There exist $c,\lambda^\ast>0$ and an exponent $d_s>0$ such that}
\begin{equation}
    \mathcal{N}(\lambda) \;\ge\; c\,\lambda^{d_s/2}
    \qquad\text{for all } 0<\lambda\le\lambda^\ast .
    \label{eq:A1}
\end{equation}
Integrating the density of states of Eq.~\eqref{eq:ds-dos}, $\rho(\lambda) \sim \lambda^{d_s/2 - 1}$, gives $\mathcal{N}(\lambda)-\mathcal{N}(0)\sim\lambda^{d_s/2}$.
(A1) retains only the lower half of this statement --- the graph must have at least this many soft modes --- and requires that its prefactor does not vanish as $N\to\infty$.

\medskip
\textbf{(A2) Resolution.}
\emph{At the site $i$ considered, there exists $C>0$ such that}
\begin{equation}
    T_i(\lambda) \;\le\; \frac{C}{\mathcal{N}(\lambda)}
    \qquad\text{for all } 0\le\lambda\le\lambda_{\max} .
    \label{eq:A2}
\end{equation}
By Eq.~\eqref{eq:T-distance}, (A2) states that the modes of energy at most $\lambda$ reproduce the step $\chi_i$ up to a squared error of order $1/\mathcal{N}(\lambda)$.
(A2) makes no reference to the form of individual eigenvectors, and it is insensitive to degeneracies and to the choice of eigenbasis.
It holds automatically on $0\le\lambda<\lambda_1$ whenever $C\ge x(1-x)$, since there $T_i=x(1-x)N$ and $\mathcal{N}=1/N$.
It holds trivially at $\lambda=\lambda_{\max}$, where $T_i=0$.
It is the only assumption that involves the labeling.
For translation-invariant graphs it can be proven [Remark~(i)].

\medskip
\textbf{(A3) Bounded spectrum.}
\emph{The spectrum has a high-energy cutoff, $\lambda_{\max} \le \Lambda$ uniformly in $N$.}

We split the integral of Eq.~\eqref{eq:layer-cake} at $\lambda^\ast$.
For $\lambda^\ast<\lambda\le\lambda_{\max}$, (A2) and the monotonicity of $\mathcal{N}$ give $T_i(\lambda)\le C/\mathcal{N}(\lambda)\le C/\mathcal{N}(\lambda^\ast)$.
(A1) at $\lambda=\lambda^\ast$ gives $\mathcal{N}(\lambda^\ast)\ge c\,\lambda^{\ast\,d_s/2}$.
By (A3), this interval has length at most $(\Lambda-\lambda^\ast)_+:=\max(\Lambda-\lambda^\ast,0)$, so it contributes at most
\begin{equation}
    R := \frac{C\,(\Lambda-\lambda^\ast)_+}{c\,\lambda^{\ast\,d_s/2}} ,
    \label{eq:R}
\end{equation}
an $N$-independent constant; the interval is empty if $\lambda_{\max}\le\lambda^\ast$.

For $0<\lambda\le\min(\lambda^\ast,\lambda_{\max})$, (A2) and (A1) give $T_i(\lambda)\le C/\mathcal{N}(\lambda)\le C/(c\,\lambda^{d_s/2})$, while property (ii) gives $T_i(\lambda)\le N/4$.
Keeping the better of the two bounds, and extending the integration to $\lambda^\ast$ if $\lambda_{\max}<\lambda^\ast$ (the integrand is non-negative, so this only increases the bound),
\begin{equation}
    \mathrm{Cut}(S_i) \;\le\; \int_0^{\lambda^\ast}\min\!\left(\frac{N}{4},\;\frac{C}{c\,\lambda^{d_s/2}}\right)d\lambda \;+\; R .
    \label{eq:cut-master}
\end{equation}
The two bounds cross at $\lambda_{\rm c}:=(4C/cN)^{2/d_s}$, below which the first is the smaller one.
Like the gap, $\lambda_{\rm c}$ scales as $N^{-2/d_s}$.
Above $\lambda_{\rm c}$ the integrand is $\propto\lambda^{-d_s/2}$.
Everything turns on whether this power is integrable at $\lambda\to0$, that is, on the sign of $1-d_s/2$.

\emph{Case $d_s<2$.}
The power is integrable at the origin, and we may drop the minimum:
\begin{equation}
    \mathrm{Cut}(S_i) \;\le\; \frac{C}{c}\int_0^{\lambda^\ast}\lambda^{-d_s/2}\,d\lambda + R
    = \frac{C}{c}\,\frac{\lambda^{\ast\,1-d_s/2}}{1-d_s/2} + R =: K ,
    \label{eq:cut-dsless2}
\end{equation}
for every $N$.
The bound is dominated by $\lambda$ close to $\lambda^\ast$ and by $R$.
The cost of a cut is then an ultraviolet quantity, set by the structure of the graph near the cut, and it does not grow with $N$.

\emph{Case $d_s=2$.}
For $N\ge4C/(c\lambda^\ast)$, so that $\lambda_{\rm c}=4C/(cN)\le\lambda^\ast$,
\begin{equation}
    \mathrm{Cut}(S_i) \;\le\; \frac{N}{4}\,\lambda_{\rm c} + \frac{C}{c}\int_{\lambda_{\rm c}}^{\lambda^\ast}\frac{d\lambda}{\lambda} + R
    = \frac{C}{c}\left[1+\ln\frac{c\lambda^\ast N}{4C}\right] + R .
    \label{eq:cut-dseq2}
\end{equation}
The integrand $C/(c\lambda)$ is scale invariant: every interval $[a,\,ba]$ with fixed ratio $b>1$ contributes the same amount, $(C/c)\ln b$, independently of $a$.
The number of such intervals between $\lambda_{\rm c}\propto 1/N$ and $\lambda^\ast$ grows as $\ln N$, which is the origin of the logarithm.

\emph{Case $d_s>2$.}
For $N\ge4C/(c\lambda^{\ast\,d_s/2})$, so that $\lambda_{\rm c}\le\lambda^\ast$, we use $\int_{\lambda_{\rm c}}^{\lambda^\ast}\lambda^{-d_s/2}\,d\lambda\le\lambda_{\rm c}^{1-d_s/2}/(d_s/2-1)$.
We also use $\frac{N}{4}\lambda_{\rm c}=\frac{C}{c}\lambda_{\rm c}^{1-d_s/2}=\frac14(4C/c)^{2/d_s}N^{1-2/d_s}$.
This gives
\begin{equation}
    \mathrm{Cut}(S_i) \;\le\; \frac{N}{4}\,\lambda_{\rm c} + \frac{C}{c}\,\frac{\lambda_{\rm c}^{\,1-d_s/2}}{d_s/2-1} + R
    = \kappa\,N^{1-2/d_s} + R ,
    \label{eq:cut-dsgreater2}
\end{equation}
where $\kappa := \frac14\left(\frac{4C}{c}\right)^{2/d_s}\frac{d_s}{d_s-2}$.
The bound is dominated by $\lambda\sim\lambda_{\rm c}\propto N^{-2/d_s}$.
The cost of a cut is then an infrared quantity, carried by the softest modes, and it grows with $N$.
Together with Eq.~\eqref{eq:fiedler}, this shows that the upper bound of Eq.~\eqref{eq:Pb-upper-bound} is tight up to the constant in the exponent.

Inserting Eqs.~\eqref{eq:cut-dsless2}--\eqref{eq:cut-dsgreater2} into Eq.~\eqref{eq:sandwich-2}, with $\lambda_{\max}\le\Lambda$,
\begin{equation}
    P_{\rm cut}(i) \;\ge\;
    \begin{cases}
        \exp\!\left[-\dfrac{K+\Lambda}{1-\overline{p}_{\max}}\right], & d_s<2,\\[12pt]
        A_-\,N^{-\theta_+}, & d_s=2,\\[6pt]
        \exp\!\left[-\dfrac{\kappa\,N^{1-2/d_s}+R+\Lambda}{1-\overline{p}_{\max}}\right], & d_s>2,
    \end{cases}
    \label{eq:Pb-lower}
\end{equation}
with
\begin{equation}
\begin{split}
    &\theta_+ := \frac{C}{c\,(1-\overline{p}_{\max})}, \\
    &A_- := \exp\!\left[-\frac{C/c+R+\Lambda}{1-\overline{p}_{\max}}\right]\left(\frac{c\lambda^\ast}{4C}\right)^{-\theta_+} .
\end{split}
\end{equation}
For $d_s<2$, the cutting probability is bounded away from zero uniformly in $N$ and in the position of the cut; the position enters only through the constant $C$ of (A2).

At $d_s=2$ the upper bound of Eq.~\eqref{eq:Pb-upper-bound-gap} is $O(1)$, so Eq.~\eqref{eq:Pb-lower} alone does not show that $P_{\rm cut}$ vanishes.
This requires the reverse of (A1) and (A2) on the soft part of the spectrum, with constants $c',C'>0$ independent of $N$:

\medskip
\textbf{(A1$'$)} \emph{$\mathcal{N}(\lambda)\le c'\lambda$ for $\lambda_1\le\lambda\le\lambda^\ast$.}

\textbf{(A2$'$)} \emph{$T_i(\lambda)\ge C'/\mathcal{N}(\lambda)$ for $\lambda_1\le\lambda\le\lambda^\ast$.}
\medskip

For $N>1/(c\lambda^\ast)$, we have $\lambda_1\le1/(cN)<\lambda^\ast$.
Dropping the non-negative contributions from outside $[\lambda_1,\lambda^\ast]$ in Eq.~\eqref{eq:layer-cake},
\begin{equation}
\begin{split}
    \mathrm{Cut}(S_i)
      &\ge \int_{\lambda_1}^{\lambda^\ast} T_i(\lambda)\,d\lambda \\
      &\ge \frac{C'}{c'} \int_{\lambda_1}^{\lambda^\ast} \frac{d\lambda}{\lambda} \\
      &= \theta_-\ln\frac{\lambda^\ast}{\lambda_1} \\
      &\ge \theta_-\ln\!\left(c\lambda^\ast N\right),
\end{split}
\label{eq:cut-marginal-lower}
\end{equation}
where $\theta_- := C'/c'$.
By Eq.~\eqref{eq:P-cut-bound}, $P_{\rm cut}(i)\le\left(c\lambda^\ast N\right)^{-\theta_-}=A_+\,N^{-\theta_-}$, with $A_+:=(c\lambda^\ast)^{-\theta_-}$.

\subsection{Summary}
\label{app:summary}
We collect Eqs.~\eqref{eq:Pb-upper-bound}, \eqref{eq:Pb-lower}, and~\eqref{eq:cut-marginal-lower}, with the same exponent $d_s$ in Eq.~\eqref{eq:ds-gap} and in (A1), and for $N$ large enough in the sense specified above:
\begin{equation}
\begin{aligned}
    &d_s<2: && e^{-\frac{K+\Lambda}{1-\overline{p}_{\max}}} \;\le\; P_{\rm cut}(i) \;\le\; 1,\\
    &d_s=2: && A_-\,N^{-\theta_+} \;\le\; P_{\rm cut}(i) \;\le\; A_+\,N^{-\theta_-},\\
    &d_s>2: && e^{-\frac{\kappa N^{1-2/d_s}+R+\Lambda}{1-\overline{p}_{\max}}} \;\le\; P_{\rm cut}(i) \;\le\; e^{-g\,x(1-x)\,N^{1-2/d_s}} .
\end{aligned}
\label{eq:Pb-bounds-summary}
\end{equation}
In short,
\begin{equation}
    P_{\rm cut} \;\sim\;
    \begin{cases}
        \mathrm{const} > 0, & d_s < 2 \;\; \text{(blocking survives)}, \\[4pt]
        N^{-\theta(p)}, & d_s = 2 \;\; \text{(marginal)}, \\[4pt]
        e^{- \mathrm{const}\cdot N^{1 - 2/d_s}}, & d_s > 2 \;\; \text{(no blocking)},
    \end{cases}
    \label{eq:Pb-dichotomy}
\end{equation}
where at the marginal point the exponent is bracketed as $\theta_-\le\theta(p)\le\theta_+$.
This identifies
\begin{equation}
    d_s^* = 2
\end{equation}
as the threshold spectral dimension separating geometries that support absolute cuts at a finite density of sites from those that do not.

\subsection{Remarks}
\label{app:remarks}

\paragraph{(i) Translation-invariant graphs.}
Suppose $\overline{p}_{a,b}=f\big((b-a)\bmod N\big)$, with $f(0)=0$ and $f(r)=f(N-r)$, as for a ring.
Then $\mathbb{L}$ is circulant.
The Fourier modes $e_k(a)=N^{-1/2}e^{2\pi\mathrm{i}ka/N}$, with $k\in\{-\lceil N/2\rceil+1,\dots,\lfloor N/2\rfloor\}$, form an orthonormal eigenbasis.
It is complex, but it spans the same eigenspaces as any real one, so it can be used in Eq.~\eqref{eq:T-def}.
The eigenvalues are $\lambda(k)=\sum_{r=1}^{N-1}f(r)\,[1-\cos(2\pi kr/N)]=\lambda(-k)$.

Assume that $\lambda(k)$ is non-decreasing in $|k|$ for $0\le|k|\le N/2$.
Then, for every $\lambda\ge0$, the modes with $\lambda(k)\le\lambda$ are those with $|k|\le K$ for some integer $K\ge0$.
As long as not all modes are included, $N\mathcal{N}(\lambda)=2K+1$; if all are included, $T_i=0$.
If $K=0$, then $T_i=\|w_i\|^2\le N/4$ and $\mathcal{N}=1/N$, so $T_i\,\mathcal{N}\le1/4$.
If $K\ge1$, the geometric sum gives, for $1\le|k|\le N/2$,
\begin{equation}
    |\langle e_k,\chi_i\rangle|^2 = \frac{\sin^2(\pi k i/N)}{N\,\sin^2(\pi k/N)} \;\le\; \frac{N}{4k^2},
    \label{eq:circ-overlap}
\end{equation}
where we used $\sin y\ge 2y/\pi$ for $0\le y\le\pi/2$.
Since $\langle e_k,w_i\rangle=\langle e_k,\chi_i\rangle$ for $k\neq0$, and $\sum_{k>K}k^{-2}\le\int_K^\infty dy/y^2=1/K$,
\begin{equation}
\begin{split}
    &T_i(\lambda) = \sum_{K<|k|\le N/2}|\langle e_k,\chi_i\rangle|^2
    \;\le\; 2\sum_{k>K}\frac{N}{4k^2} \;\le\; \frac{N}{2K},
    \\
    &T_i(\lambda)\,\mathcal{N}(\lambda)\;\le\;\frac{2K+1}{2K}\;\le\;\frac32 .
\end{split}
\label{eq:circ-tail}
\end{equation}
Hence (A2) holds with $C=3/2$ at every site and for every $N$, whatever the range of $f$, and in particular also when $d_s>2$.
(A2) thus expresses compatibility between labeling and connectivity and is independent of $d_s$; which regime applies is decided by (A1).

Monotonicity of $\lambda(k)$ holds, e.g., for the nearest-neighbor ring, and can be checked directly for the power-law ring of Sec.~\ref{sec:lr_graph}.
It is not a technicality: for circulants with an irregular, non-decaying $f$ the dispersion is not monotone and (A2) can fail.

\section{Asymptotic behavior of the cutting probability}
\label{app:asympt}
We study the asymptotic behavior of $P_{\rm cut}$ as $N \to \infty$ for the two example graphs; note that increasing the system size also increases the circuit depth.

\subsection{Long-range graph}
In this graph, $p_{a,b} = p_r = r^{-\alpha}$, where $r = |a - b|$ in the 1D embedding.
Noting that for a fixed distance $r$, $r+1$ bonds cross a cut, Eq.~\eqref{eq:ln_Pb} collapses to
\begin{equation}
    -\ln P_{\rm cut}
    = -\sum_{r=1}^{N-1} (r+1)
        \ln\!\left( 1 - \frac{1-p}{r^\alpha} \right),
    \label{eq:long_range_lnPcut}
\end{equation}
where the $r=0$ term drops out since $p_{a,a}=0$.

For $p = 1$ it is immediate that $P_{\rm cut} = 1$. For $p < 1$, a divergent sum implies $P_{\rm cut} = 0$ and a convergent sum a finite $P_{\rm cut}$.
Using $\ln(1 - x) \sim -x$ yields
\begin{equation}
    -\ln P_{\rm cut} \sim \sum_{r=1}^{N-1} (r+1)\,\frac{1-p}{r^\alpha}
          \sim (1-p)\sum_{r=1}^{N-1} \frac{1}{r^{\alpha - 1}}.
\end{equation}
This is a generalized harmonic series, and for $N \to \infty$,
\begin{align}
    \sum_{r=1}^{\infty}\frac{1}{r^{\alpha-1}}
    =
    \begin{cases}
        \infty, & \alpha \le 2, \\
        \zeta(\alpha - 1), & \alpha > 2,
    \end{cases}
\end{align}
while for large but finite $N$ the integral approximation gives
\begin{align}
    \sum_{r=1}^{N-1}\frac{1}{r^{\alpha-1}}
    \approx \int_1^{N} r^{1-\alpha}\, dr =
    \begin{cases}
        \dfrac{N^{2-\alpha} - 1}{2 - \alpha}, & \alpha \neq 2, \\[4pt]
        \ln N, & \alpha = 2 .
    \end{cases}
\end{align}
Collecting the two,
\begin{align}
    -\ln P_{\rm cut}
    \sim
    \begin{cases}
        N^{2-\alpha}, & 0<\alpha<2, \\
        \ln N, & \alpha =2, \\
        \zeta(\alpha-1), & \alpha>2,
    \end{cases}
\end{align}
and therefore
\begin{align}
    P_{\rm cut}
    \to
    \begin{cases}
        0, & 0<\alpha\le 2, \\
        \text{finite}<1, & \alpha>2,
    \end{cases}
    \qquad N \to \infty .
\end{align}

The asymptotics of $P_{\rm cut}$ constrain the phase structure.
For $P_{\rm cut} > 0$ as $N \to \infty$ we have $p_c \to 0$: for any $p>0$ there is a nonzero probability that a single qubit splits the chain in two, preventing thermalization.
For $P_{\rm cut} \to 0$ no direct conclusion follows.
However, for $\alpha < 1$ the mean connectivity per qubit diverges and we expect thermalization for any $p<1$, whereas in the intermediate regime $1 < \alpha < 2$ a critical value $p_c(\alpha)$ emerges from the competition between blocking gates and long-range correlations.
This is the standard percolation intuition: if the effective number of long bonds per site diverges, percolation is guaranteed for any $p<1$ ($p_c \to 1$); if long bonds are sparse, percolation may require $p \to 0$.

\subsection{PWR2 graph}
In this graph, $p_{a,b} = p_r = r^{-\alpha}$ for $r=2^n$ and $0$ otherwise, with $n=0,1,\dots,n_\mathrm{max}$, and Eq.~\eqref{eq:ln_Pb} becomes
\begin{equation}
    -\ln P_{\rm cut}
    = -\sum_{n=0}^{n_\mathrm{max}}
        (2^n + 1)
        \ln\!\left( 1 - \frac{1-p}{2^{n\alpha}} \right),
    \label{eq:pwr2_lnPcut}
\end{equation}
where $n_\mathrm{max} = \lfloor \log_2(N-1) \rfloor$.
For $p=1$, again $P_{\rm cut}=1$.
For $p<1$, using $\ln(1-x)\sim -x$,
\begin{equation}
\ln P_{\rm cut} \sim -\sum_{n=0}^{n_\mathrm{max}}
        (2^n + 1) \frac{1-p}{2^{n\alpha}}
          \sim -(1-p)\sum_{n=0}^{n_\mathrm{max}} \frac{1}{2^{n(\alpha - 1)}} ,
\end{equation}
a geometric series with ratio $a = 2^{1-\alpha}$, which converges iff $\alpha > 1$.
Therefore
\begin{align}
    P_{\rm cut}
    \to
    \begin{cases}
        0, & 0<\alpha\le 1, \\
        \text{finite}<1, & \alpha>1,
    \end{cases}
    \qquad N \to \infty ,
\end{align}
so the critical value is $\alpha = 1$ (corresponding to $d_s = 2$), and the transition window is $0 < \alpha < 1$.

\section{Stabilizer formalism \& canonical form}
\label{app:stabilizer_formalism}
In the stabilizer formalism, a quantum state $\ket{\psi}$ is represented by its \textit{stabilizer group}: the set of unitaries $U$ for which $U\ket{\psi} = \ket{\psi}$, i.e., for which $\ket{\psi}$ is an eigenvector with eigenvalue 1.
If $U$ and $V$ stabilize $\ket{\psi}$, so do $UV$ and $U^{-1}$; hence $\text{Stab}(\ket{\psi})$ forms a group.
For an $N$-qubit state $\ket{\psi}$ obtained from $\ket{0}^{\otimes N}$ by Clifford gates (generated by CNOT, Hadamard, and the phase gate), $\ket{\psi}$ is uniquely determined by $\mathcal{S}(\ket{\psi}) = \text{Stab}(\ket{\psi}) \cap \mathcal{P}_N$, the subgroup of its stabilizers lying in the Pauli group $\mathcal{P}_N$~\cite{gottesman1998heisenberg,aaronson2004improved}.

The stabilizer group can be represented by a generating set $\mathcal{S} = \langle g_1, \dots, g_N \rangle \subset \mathcal{P}_N$, of size $N = \log_2 2^N$, with $-\id \notin \mathcal{S}$. To linearize the group structure, each generator is written in the binary \emph{symplectic representation}~\cite{calderbank1998QEC,gottesman1997thesis,dehaene2003clifford}, which encodes Pauli strings as vectors over $\mathbb{F}_2$:
\begin{equation}
    g_a \;=\; i^{\delta_a} \prod_{j=1}^{N} X_j^{\,x_{aj}} Z_j^{\,z_{aj}} ,
    \qquad x_{aj}, z_{aj} \in \mathbb{F}_2 ,\quad \delta_a \in \{0,2\},
\end{equation}
where, as usual, $X_j Z_j$ stands for $Y_j$, and $\delta_a \in \{0,2\}$ fixes the sign of $g_a$ to $\pm 1$, as $g_a$ has to be Hermitian with eigenvalue $+1$.
The state is then stored as the $N \times 2N$ binary \emph{tableau}
\begin{equation}
    T \;=\; \left( \, \mathsf{X} \; \middle| \; \mathsf{Z} \, \right),
    \qquad \mathsf{X}, \mathsf{Z} \in \mathbb{F}_2^{N \times N},
\end{equation}
supplemented by the phase bits $\delta_a$, which play no role in what follows; $\mathbb{F}_2^{N \times N}$ are the binary matrices under mod-2 arithmetic.

Row operations on $T$ (multiplication of generators) leave the state invariant, and single-qubit Cliffords leave every entanglement measure invariant.
Both freedoms can be used to bring $T$ into a \textit{canonical form}~\cite{vandennest2004graph}, achieved by sweeping over columns $i = 1, \dots, N$, and applying the following steps:
\begin{enumerate}
    \item \emph{Pivot.} If $\mathsf{X}_{ii} = 0$, swap row $i$ with a row $j>i$
          having $\mathsf{X}_{ji} = 1$. If the whole column vanishes, apply a
          Hadamard on qubit $i$ --- which exchanges the columns
          $\mathsf{X}_{:,i} \leftrightarrow \mathsf{Z}_{:,i}$ --- and pivot again.
    \item \emph{Eliminate.} Add row $i$ to every row $j \neq i$ with
          $\mathsf{X}_{ji} = 1$, so that column $i$ of $\mathsf{X}$ is cleared.
\end{enumerate}
At the end $\mathsf{X} = \id$ and the generators read
$g_i = X_i \prod_{j} Z_j^{\Theta_{ij}}$ with $\Theta = \mathsf{Z}$.
The symplectic condition now forces $\Theta = \Theta^{T}$, so $\Theta$ is the
adjacency matrix of an undirected graph $G_\psi$ on the $N$ qubits.
A residual diagonal entry $\Theta_{ii} = 1$ means $g_i$ contains $Y_i$ rather
than $X_i$ and is removed by a local phase gate $S$ on qubit $i$.
Hence
\begin{equation}
    \ket{\psi} \;=\; \Big( \bigotimes_{i=1}^{N} C_i \Big) \ket{G_\psi} ,
    \qquad
    \ket{G_\psi} = \prod_{(i,j) \in E(G_\psi)} \mathrm{CZ}_{ij} \ket{+}^{\otimes N},
    \label{eq:graph_state_}
\end{equation}
with $C_i \in \{ \id, H, S, \dots \}$ single-qubit Cliffords.

Notice the distinction between two graphs: the \emph{interaction graph} of Sec.~\ref{sec:model} specifies where the gates act and is fixed by the realization; the \emph{entanglement graph} $G_\psi$
defined by Eq.~\eqref{eq:graph_state_} is a property of the evolved state and is read
off the tableau.
The former is the input of the dynamics, the latter its output.

\section{Numerical simulations of the Clifford dynamics}
\label{app:details_cliff}
Circuit simulations in this work are performed within the stabilizer formalism (see App.~\ref{app:stabilizer_formalism}).
For each choice of the parameters $(N, p, \alpha)$ we generate $R$ independent realizations.
For each realization, we first sample the interaction graph and the gates on its edges; compose the gates into the Floquet unitary $U$; evolve the initial state $\ket{0}^{\otimes N}$ until it reaches the steady state; and sample the observables over a window of Floquet periods in the steady state.
We describe in detail these steps below; the full implementation is available with the code accompanying this work~\cite{code_repo}.

Up to phases, a Clifford unitary is fully specified by how it transforms the $2N$ single-qubit Paulis $X_j$ and $Z_j$ into other Pauli strings.
In the binary representation of App.~\ref{app:stabilizer_formalism}, where a Pauli string is a row vector $(\mathbf{x}\,|\,\mathbf{z}) \in \mathbb{F}_2^{2N}$, a Clifford $U$ is therefore represented by a $2N \times 2N$ binary symplectic matrix $M_U$, whose columns are the images of $X_1, \dots, X_N, Z_1, \dots, Z_N$.
The conjugation of all the stabilizer generators, $g_a \mapsto U g_a U^\dagger$, then acts on the tableau as a single matrix product,
\begin{equation}
    T \;\mapsto\; T \, M_U^{T} \pmod{2} .
    \label{eq:tableau_update}
\end{equation}
None of the entanglement quantities considered here depends on the phase bits $\delta_a$; we therefore do not track them, and the dynamics reduces entirely to linear algebra over $\mathbb{F}_2$.
A two-qubit gate on the sites $(i,j)$ acts only on the four components $(x_i, x_j, z_i, z_j)$, and is represented by a $4 \times 4$ symplectic matrix $S$, embedded in $M_U$ acting as the identity on all the other components.
For the gates introduced in Sec.~\ref{sec:model},
\begin{equation}
    M_{\rm B} =
    \begin{pmatrix}
        1 & 0 & 0 & 0 \\
        0 & 1 & 0 & 0 \\
        0 & 1 & 1 & 0 \\
        1 & 0 & 0 & 1
    \end{pmatrix},
    \qquad
    M_{\rm NB} =
    \begin{pmatrix}
        0 & 1 & 0 & 1 \\
        1 & 1 & 0 & 1 \\
        1 & 1 & 1 & 0 \\
        1 & 0 & 1 & 0
    \end{pmatrix},
    \label{eq:gate_symplectic}
\end{equation}
whose four columns are the images of $XI$, $IX$, $ZI$, $IZ$ listed in Eq.~\eqref{eq:BNBdefinition}.

We arrange $N$ qubits on a ring with periodic boundary conditions, labeled $i = 1, \dots, N$.
For every site $i$ and every allowed distance $r$, the edge $(i, \, i+r)$, with site labels understood modulo $N$, is realized independently with probability $p_r = r^{-\alpha}$, where the allowed distances are $r = 1, \dots, N/4$ for the LR graph and $r = 2^n$, with $n = 0, 1, \dots,\lfloor\log_2(N/4)\rfloor$, for the PWR2 graph.
Every unordered pair of sites is considered exactly once, and, since $p_1 = 1$, the nearest-neighbor ring is always present.
Each realized edge then independently hosts a B gate with probability $p$ and an NB gate with probability $1-p$.
The cutoff $r \le N/4$ on the gate range, introduced to mitigate finite-size effects, matches the support of the QCMI, in that no single gate acts on both $A$ and $C$.
Since the cutoff scales with $N$, it only modifies the prefactor of the spectral gap $\lambda_1$.

To compose the Floquet unitary $U$, the gates are applied one after the other sweeping the sites in increasing order $i = 1, \dots, N$ and, for each $i$, the edges $(i, i+r)$ in increasing order of $r$, with site $i$ as the first and site $i+r$ as the second qubit of the gate.
Therefore, $M_U = M^{(G)} \cdots M^{(2)} M^{(1)}$, where $M^{(k)}$ is the embedded matrix of the $k$-th gate and $G$ the total number of gates.
Each factor modifies only four rows of the product, so that $M_U$ is assembled in $O(N^2)$ operations.
While $M_{\rm B}$ is symmetric under the exchange of the two qubits, $M_{\rm NB}$ is not; we verified on small systems that drawing the orientation of every gate at random leaves the steady-state entanglement and QCMI unchanged.

At the largest system sizes, $N = 1024$--$4096$, the construction of $M_U$ is the computational bottleneck.
A direct implementation would loop, for every site, over all the $N/4$ allowed distances, i.e., $O(N^3)$ operations per realization --- $N^2/4$ sequential gate applications of cost $O(N)$ each.
To avoid this, we note that the number of edges $(i, i+r)$ from site $i$ is a sum of independent Bernoulli variables, with mean and variance
\begin{equation}
    \mu = \sum_r p_r , \qquad \sigma^2 = \sum_r p_r \left( 1 - p_r \right) ,
    \label{eq:edge_number_moments}
\end{equation}
which depend only on $\alpha$ and $N$ and converge to finite values as $N \to \infty$ (at $1<\alpha<2$ for the LR graph and $0 < \alpha < 1$ for the PWR2 graph).
For each site, we first draw the Bernoulli variables for all the allowed distances at once and then store the realized edges, in increasing order of $r$, into a buffer of fixed length
\begin{equation}
    K = \left\lceil \mu + 5 \sigma \right\rceil ,
    \label{eq:K_cap}
\end{equation}
padded with void entries that act as the identity.
The sequential loop then runs over these $K$ entries only, so that $M_U$ is assembled with $NK$ sequential gate applications and $O(K N^2)$ operations.
For the LR graph, $K$ saturates to $\lceil \zeta(\alpha) + 5 \sqrt{\zeta(\alpha) - \zeta(2\alpha)} \, \rceil$ as $N \to \infty$, and ranges between $6$ and $8$ for all the parameters of Table~\ref{tab:datasets}; at $N = 4096$, this replaces the loop over $N/4 = 1024$ distances by one over at most $8$ entries, reducing the number of sequential steps by about two orders of magnitude.
For the PWR2 graph, $K$ is further capped at the number $\log_2(N/4) + 1$ of allowed distances, and the gain is modest, since this graph is sparse by construction.
The compression is not strictly exact, as a site with more than $K$ connections keeps only its $K$ shortest edges.
For the parameters used here, this happens with probability below $3 \times 10^{-5}$ per site for LR and below $10^{-6}$ for PWR2, so that at most a fraction $\sim 10^{-5}$ of the edges is discarded, a negligible modification of the edge distribution.

The ordering of the gates in the Floquet unitary $U$ is not unique, and for a given realization different orderings define different unitaries.
However, the observables are ensemble and time averages in the steady state, which are controlled by the B/NB structure of the interaction graph (Sec.~\ref{sec:stab_formalism_percolation} and App.~\ref{app:BNB_gates}) rather than by the order in which the gates are applied.
The choice of ordering therefore leaves our results unchanged, and only affects the transient dynamics, i.e., the number of Floquet periods needed to reach the steady state.
As we show next, the cost of the time evolution grows only logarithmically with the number of periods, so that this is irrelevant for our purposes.

All simulations start from the product state $\ket{0}^{\otimes N}$, stabilized by $Z_1, \dots, Z_N$, i.e., $T(0) = (\,0 \,|\, \id\,)$.
After $t$ Floquet periods, Eq.~\eqref{eq:tableau_update} gives $T(t) = T(0) \, \big( M_U^{t} \big)^{T}$.
Since the same $U$ acts at every period, for $t = 2^k$ the power $M_U^{t}$ is obtained by $k$ successive squarings, $M_U \to M_U^2 \to M_U^4 \to \cdots$, each of which is a single $2N \times 2N$ matrix product.
Reaching time $t$ thus costs $O(N^3 \log_2 t)$ operations, instead of the $O(N^3\, t)$ of a period-by-period evolution.
Empirically, for all the system sizes and parameters considered, the transient dynamics lasts fewer than $N$ Floquet periods.
We evolve each realization for $t_{\rm eq} = 2N$ periods, i.e., $\log_2 N + 1$ squarings, and then apply $U$ one period at a time via Eq.~\eqref{eq:tableau_update} for further $t_{\rm meas}$ periods, over which the observables are sampled.

The entanglement entropy of a region $A$ in a stabilizer state is $S_A = |A| - \dim \mathcal{S}_A$, where $\mathcal{S}_A$ is the subgroup of stabilizers supported entirely within $A$~\cite{fattal2004entanglement}.
Rather than evaluating an $\mathbb{F}_2$ rank for each region, we bring the tableau into the \emph{clipped gauge}~\cite{nahum2017entanglementgrowth,li2019measurement}, in which the entropy of any contiguous region follows from the supports of the generators alone.
Let $l_a$ and $r_a$ denote the leftmost and rightmost sites on which $g_a$ acts nontrivially.
The tableau is in the clipped gauge if every site hosts exactly two endpoints~\cite{nahum2017entanglementgrowth}.
The gauge is reached by row operations only through two sweeps of Gaussian elimination, which leave the state invariant; the cost is $O(N^3)$.
In the clipped gauge, the product of any subset of generators acts nontrivially on the leftmost left endpoint and on the rightmost right endpoint within the subset.
The generators with both endpoints inside a contiguous region $A$ thus form a basis of $\mathcal{S}_A$, and counting the $2|A|$ endpoints lying in $A$ yields
\begin{equation}
    S_A \;=\; \frac{1}{2} \, \#\big\{ a \,:\, \text{only one of } l_a, r_a \text{ lies in } A \big\} ,
    \label{eq:clipped_entropy}
\end{equation}
which for $A = \{1, \dots, x\}$ reduces to the familiar $S(x) = \frac{1}{2} \#\{ a : l_a \le x < r_a \}$~\cite{nahum2017entanglementgrowth}.
Once the gauge is fixed, Eq.~\eqref{eq:clipped_entropy} gives the entropy of any contiguous region in $O(N)$ operations.

Our main observable is the quantum conditional mutual information (QCMI)~\cite{otsuka2025qcmi}
\begin{equation}
    I(A:C|B) = S_{A \cup B} + S_{B \cup C} - S_B - S_{A \cup B \cup C},
\end{equation}
for the partition of the ring into four consecutive quarters $A = \{1, \dots, N/4\}$, $B = \{N/4+1, \dots, N/2\}$, $C = \{N/2+1, \dots, 3N/4\}$, and $D = \{3N/4+1, \dots, N\}$ [inset of Fig.~\ref{fig:phase_diagram}(a)].
All the regions entering $I(A:C|B)$ are contiguous and do not wrap around the ring, so that the four entropies follow directly from Eq.~\eqref{eq:clipped_entropy}.
We also record the half-chain entanglement entropy $S_{N/2}$, which gives a qualitative picture of the crossover from area to volume law, but does not enter the finite-size scaling analysis.
For each realization $\mu = 1, \dots, R$, the QCMI is averaged over the measurement window,
\begin{equation}
    \bar{I}_\mu \;=\; \frac{1}{t_{\rm meas}} \sum_{t = t_{\rm eq}+1}^{t_{\rm eq}+t_{\rm meas}} I_\mu(t) ,
    \label{eq:time_avg_qcmi}
\end{equation}
which suppresses the temporal fluctuations within a realization.
The time averages $\bar{I}_\mu$, one per realization, are the raw data of the finite-size scaling analysis of App.~\ref{app:fss}; the datasets are listed in Table~\ref{tab:datasets}.

The simulations are implemented in JAX~\cite{jax2018github}.
The construction of $U$, the time evolution, and the computation of the observables are vectorized over batches of realizations, which are distributed over $4$ to $16$ NVIDIA A100 GPUs, depending on the system size.
Binary matrices are stored as $8$-bit integers, and products over $\mathbb{F}_2$ are computed as ordinary integer matrix products followed by a reduction modulo $2$; the wrap-around modulo $256$ of the $8$-bit arithmetic preserves the parity.
Per realization, the construction of $U$ requires $O(K N^2)$ operations, the equilibration $O(N^3 \log_2 N)$, and the measurement window $O(N^3\, t_{\rm meas})$, where each period requires one update of Eq.~\eqref{eq:tableau_update} and one gauge fixing.
The memory required per realization is $O(N^2)$, which sets the number of realizations that can be processed simultaneously on a single GPU.

\begin{table*}[t]
    \caption{Datasets used in the finite-size scaling analysis. For each spectral dimension $d_s$, the LR and PWR2 exponents are related by $\alpha_{\rm PWR2} = \alpha_{\rm LR} - 1$. For each family we list the system sizes $N$, the range of blocking probabilities $p$, the number $n_p$ of $p$ values per size, and the number $R$ of realizations per $(N, p)$, for the data entering the collapses, i.e., all the points of the fitted windows (Figs.~\ref{fig:fss_lr} and~\ref{fig:fss_pwr2}); ranges are over the sizes and the points.}
    \label{tab:datasets}
    \begin{ruledtabular}
    \begin{tabular}{c ccccc ccccc}
        & \multicolumn{5}{c}{LR} & \multicolumn{5}{c}{PWR2} \\
        \cline{2-6} \cline{7-11}
        $d_s$ & $\alpha$ & $N$ & $p$ & $n_p$ & $R$ & $\alpha$ & $N$ & $p$ & $n_p$ & $R$ \\
        \hline
        $40/19$ & $1.95$ & $128$--$2048$ & $0.536$--$0.716$ & $90$--$91$ & $6.6\times 10^{4}$--$1.0\times 10^{6}$ & \multicolumn{5}{c}{---} \\
        $20/9$ & $1.9$ & $128$--$2048$ & $0.620$--$0.820$ & $73$--$101$ & $6.6\times 10^{4}$--$1.0\times 10^{6}$ & $0.9$ & $128$--$2048$ & $0.740$--$0.918$ & $90$ & $1.6\times 10^{4}$--$5.2\times 10^{5}$ \\
        $5/2$ & $1.8$ & $128$--$2048$ & $0.736$--$0.914$ & $69$--$90$ & $6.6\times 10^{4}$--$1.0\times 10^{6}$ & $0.8$ & $128$--$2048$ & $0.834$--$0.954$ & $60$--$61$ & $1.6\times 10^{4}$--$5.2\times 10^{5}$ \\
        $20/7$ & $1.7$ & $128$--$2048$ & $0.839$--$0.938$ & $50$--$51$ & $3.3\times 10^{4}$--$1.0\times 10^{6}$ & $0.7$ & $128$--$2048$ & $0.892$--$0.972$ & $40$--$41$ & $1.6\times 10^{4}$--$5.2\times 10^{5}$ \\
        $10/3$ & $1.6$ & $128$--$4096$ & $0.900$--$0.958$ & $30$ & $3.3\times 10^{4}$--$5.2\times 10^{5}$ & $0.6$ & $128$--$2048$ & $0.938$--$0.978$ & $20$--$21$ & $1.6\times 10^{4}$--$5.2\times 10^{5}$ \\
    \end{tabular}
    \end{ruledtabular}
\end{table*}

\section{Details of the FSS analysis}
\label{app:fss}
\begin{figure*}[p]
    \centering
    \includegraphics[width=\textwidth]{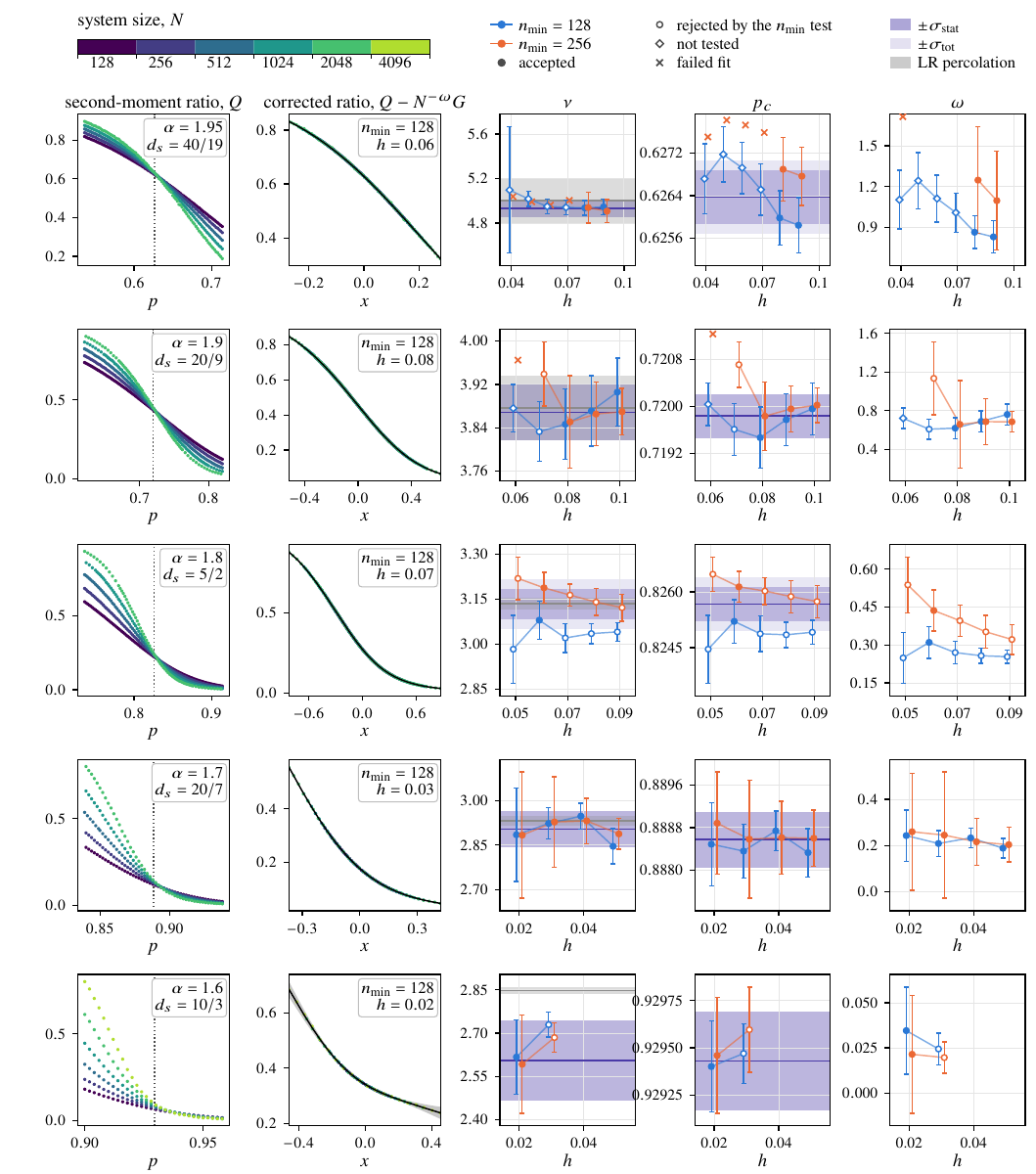}
    \caption{\textbf{Finite-size scaling analysis of the long-range graph}, one row per spectral dimension, from $d_s = 40/19$ ($\alpha = 1.95$, top) to $d_s = 10/3$ ($\alpha = 1.6$, bottom).
    First column: second-moment ratio $Q$ versus $p$ for all the sizes and points of the largest window used; the dotted line marks the final $p_c$ of Table~\ref{tab:fss_results}.
    Second column: corrected ratio $Q - N^{-\omega} G$ versus the scaling variable $x$ of Eq.~\eqref{eq:fss_model}, for the configuration $(n_{\min}, h)$ indicated. The black solid line is the master curve $F$ and its $1\sigma$ band (gray).
    Last three columns: $\nu$, $p_c$, and $\omega$ for every configuration of the grid versus the window half-width $h$, for $n_{\min} = 128$ (blue) and $n_{\min} = 256$ (orange).
    Filled circles are accepted, open circles rejected by the $n_{\min}$-consistency test, diamonds could not be tested, and crosses are failed fits; error bars are bootstrap standard deviations.
    Purple bands: final value $\pm \sigma_{\rm stat}$ (dark) and $\pm \sigma_{\rm tot}$ (light); gray band: 1D long-range percolation~\cite{gori2017onedimensional} at the same $d_s$.}
    \label{fig:fss_lr}
\end{figure*}
\begin{figure*}[p]
    \centering
    \includegraphics[width=\textwidth]{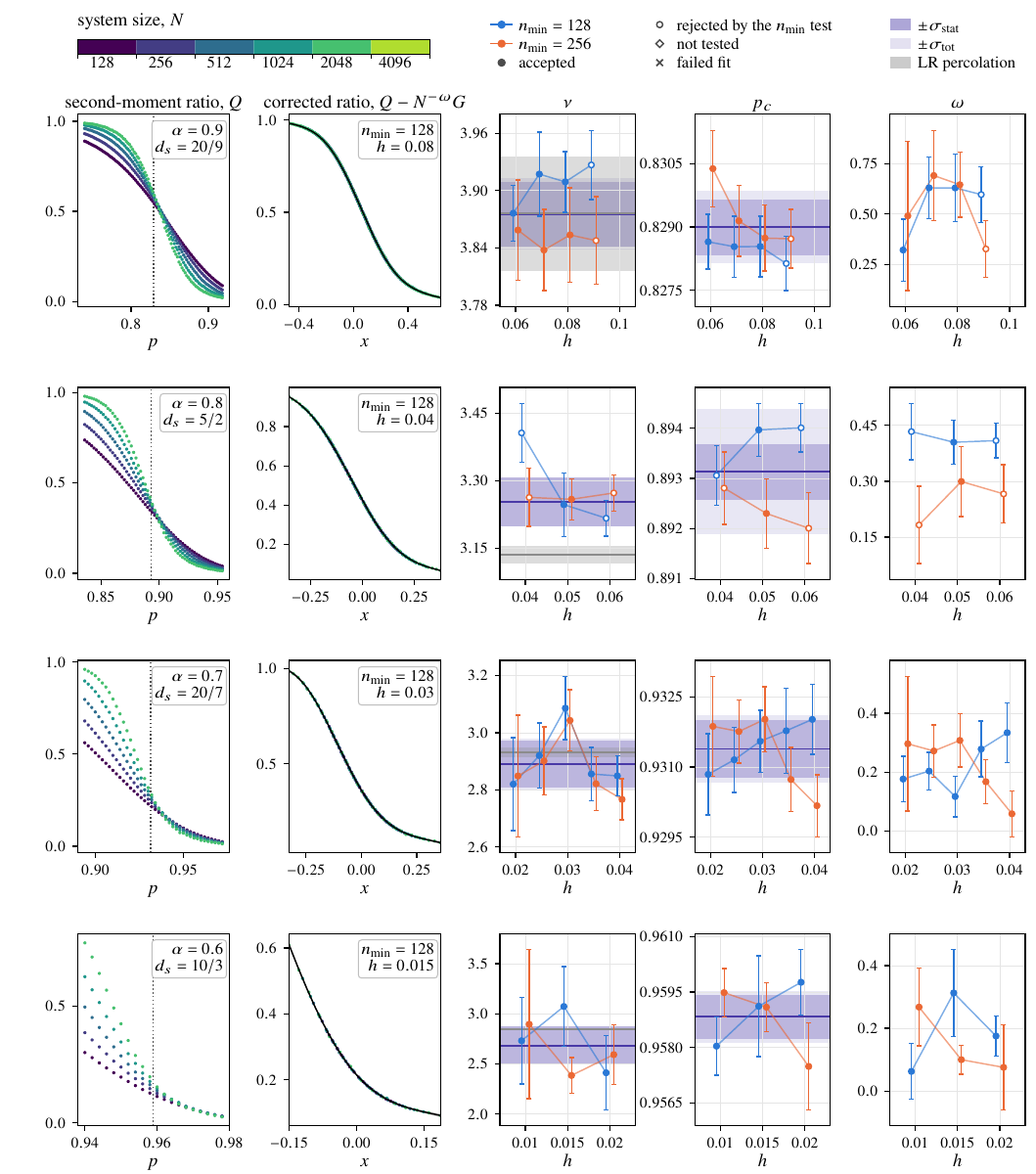}
    \caption{\textbf{Finite-size scaling analysis of the PWR2 graph}, from $d_s = 20/9$ ($\alpha = 0.9$, top) to $d_s = 10/3$ ($\alpha = 0.6$, bottom); the panels are the same as in Fig.~\ref{fig:fss_lr}, except that the corrected ratio of the second column also subtracts the second correction, $Q - N^{-\omega} G - N^{-2\omega} H$.}
    \label{fig:fss_pwr2}
\end{figure*}
\begin{table*}[t]
    \caption{Results of the finite-size scaling analysis. For each graph and spectral dimension we list the window half-widths $h$ and system sizes $N$ of the accepted configurations, their number $M$, the critical point $p_c$ with its total error, and $\nu$ with its statistical and systematic errors [Eq.~\eqref{eq:syst}]. The last two columns compare $1/\nu$ (total error) with 1D long-range percolation at the same $d_s$~\cite{gori2017onedimensional}.}
    \label{tab:fss_results}
    \begin{ruledtabular}
    \begin{tabular}{cc ccc ccc c}
        $d_s$ & $\alpha$ & $h$ & $N$ & $M$ & $p_c$ & $\nu \pm \sigma_{\rm stat} \pm \sigma_{\rm syst}$ & $1/\nu$ & $1/\nu$ (LR percolation) \\
        \hline
        \multicolumn{9}{c}{long-range} \\
        \hline
        $40/19$ & $1.95$ & $0.08$--$0.09$ & $128$--$2048$ & $4$ & $0.6264 \pm 0.0007$ & $4.932 \pm 0.083 \pm 0.000$ & $0.203 \pm 0.003$ & $0.200 \pm 0.008$ \\
        $20/9$ & $1.9$ & $0.08$--$0.1$ & $128$--$2048$ & $6$ & $0.7198 \pm 0.0004$ & $3.868 \pm 0.051 \pm 0.000$ & $0.259 \pm 0.003$ & $0.258 \pm 0.004$ \\
        $5/2$ & $1.8$ & $0.06$ & $128$--$2048$ & $2$ & $0.8257 \pm 0.0007$ & $3.133 \pm 0.050 \pm 0.065$ & $0.319 \pm 0.008$ & $0.319 \pm 0.002$ \\
        $20/7$ & $1.7$ & $0.02$--$0.05$ & $128$--$2048$ & $8$ & $0.8886 \pm 0.0005$ & $2.904 \pm 0.062 \pm 0.000$ & $0.344 \pm 0.007$ & $0.341 \pm 0.002$ \\
        $10/3$ & $1.6$ & $0.02$ & $128$--$4096$ & $2$ & $0.9294 \pm 0.0003$ & $2.605 \pm 0.139 \pm 0.000$ & $0.384 \pm 0.020$ & $0.351 \pm 0.002$ \\
        \hline
        \multicolumn{9}{c}{PWR2} \\
        \hline
        $20/9$ & $0.9$ & $0.06$--$0.08$ & $128$--$2048$ & $6$ & $0.8290 \pm 0.0009$ & $3.875 \pm 0.034 \pm 0.016$ & $0.258 \pm 0.002$ & $0.258 \pm 0.004$ \\
        $5/2$ & $0.8$ & $0.05$ & $128$--$2048$ & $2$ & $0.8931 \pm 0.0012$ & $3.252 \pm 0.054 \pm 0.000$ & $0.307 \pm 0.005$ & $0.319 \pm 0.002$ \\
        $20/7$ & $0.7$ & $0.02$--$0.04$ & $128$--$2048$ & $10$ & $0.9314 \pm 0.0007$ & $2.891 \pm 0.082 \pm 0.034$ & $0.346 \pm 0.011$ & $0.341 \pm 0.002$ \\
        $10/3$ & $0.6$ & $0.01$--$0.02$ & $128$--$2048$ & $6$ & $0.9588 \pm 0.0007$ & $2.682 \pm 0.191 \pm 0.000$ & $0.373 \pm 0.027$ & $0.351 \pm 0.002$ \\
    \end{tabular}
    \end{ruledtabular}
\end{table*}
Finite-size scaling (FSS) extracts the properties of a continuous phase transition from data at finite system sizes.
Close to the critical point $p_c$, the correlation length diverges as $\xi \sim |p - p_c|^{-\nu}$, which defines the correlation-length exponent $\nu$.
This divergence is cut off once $\xi$ becomes comparable with the system size $N$, so that, close to $p_c$, a dimensionless observable $O$ depends on $p$ and $N$ only through the ratio $\xi/N$,
\begin{equation}
    O(p, N) = F(x) ,
    \qquad
    x = (p - p_c) \, N^{1/\nu} ,
    \label{eq:fss_standard}
\end{equation}
up to corrections that vanish as $N \to \infty$.
The curves $O(p)$ of different sizes then cross at $p_c$, where $x = 0$, and collapse onto the single master curve $F$ when plotted against $x$; the values of $p_c$ and $\nu$ that produce the collapse estimate the critical point and the critical exponent.

The QCMI we sample from the circuit dynamics does not follow this form.
The curves $\langle I \rangle(p)/N$ of different sizes cross, but the crossing drifts with $N$, and the value of $\langle I \rangle$ at the intersection grows only approximately linearly with $N$.
A scaling form for $\langle I \rangle$ thus requires an $N$-dependent prefactor, $\langle I \rangle = N^{y} F(x)$, with an exponent $y$ that is not known a priori and that would have to be fitted together with $p_c$ and $\nu$.
We therefore collapse a dimensionless ratio of moments, in the spirit of the Binder cumulant~\cite{binder1981finite},
\begin{equation}
    Q = \frac{\langle I \rangle^2}{\langle I^2 \rangle} ,
    \label{eq:Q_def}
\end{equation}
in which such a prefactor cancels.
The ratio $Q$ approaches $1$ in the volume-law phase, where $I$ is self-averaging, and $0$ in the area-law phase, where $I$ vanishes in most realizations.

The moments in Eq.~\eqref{eq:Q_def} are disorder averages over the $R$ realizations of App.~\ref{app:details_cliff}, $m_k = R^{-1} \sum_{\mu} \bar{I}_\mu^{\,k}$, where $\bar{I}_\mu$ is the time average of Eq.~\eqref{eq:time_avg_qcmi}, so that $Q = m_1^2 / m_2$.
Different realizations are independent, while successive periods within one realization are correlated; statistical errors are thus estimated by treating each $\bar{I}_\mu$ as one independent sample, and $Q$ is assigned the delta-method standard error
\begin{equation}
\begin{split}
    \sigma_Q^2 = \frac{1}{R} \Big[ & d_1^2 \big( m_2 - m_1^2 \big) + d_2^2 \big( m_4 - m_2^2 \big) \\
    & + 2 d_1 d_2 \big( m_3 - m_1 m_2 \big) \Big] ,
\end{split}
    \label{eq:Q_error}
\end{equation}
with $d_1 = \partial Q / \partial m_1 = 2 m_1 / m_2$ and $d_2 = \partial Q / \partial m_2 = - m_1^2 / m_2^2$.
The resulting pairs $(Q, \sigma_Q)$ at each $(N, p)$ of Table~\ref{tab:datasets} are the input of the analysis below.

For the accessible sizes $N\in[64,4096]$, the scaling form in Eq.~\eqref{eq:fss_standard} does not describe $Q$ either, owing to finite-size effects.
We therefore fit the corrected ansatz of Eq.~\eqref{eq:fss_ansatz}, which for the data points $i$, at size $N_i$ and blocking probability $p_i$, reads
\begin{equation}
\begin{split}
    Q_i &= F(x_i) + N_i^{-\omega} G(x_i) + \epsilon_i , \\
    x_i &= \big[ (p_i - p_c) + b \, (p_i - p_c)^2 \big] N_i^{1/\nu} .
\end{split}
    \label{eq:fss_model}
\end{equation}
The term $N^{-\omega} G(x)$ is the leading correction to scaling, with exponent $\omega > 0$ and amplitude function $G$.
With respect to Eq.~\eqref{eq:fss_standard}, the scaling variable also includes the leading nonlinear term $b$ of the scaling field, which accounts for the curvature of $Q$ within windows that extend away from $p_c$.
The noise $\epsilon_i$ is Gaussian with variance $\sigma_{Q,i}^2 + \theta_2^2$, where $\theta_2$ absorbs any scatter in excess of the statistical errors.
Rather than parameterizing them, the master curve $F$ and the correction $G$ are treated as independent Gaussian processes with squared-exponential kernels~\cite{harada2011bayesian,harada2015corrections}, so that the data, centered on their mean, are jointly Gaussian with covariance
\begin{equation}
    C_{ij} = \Big[ \theta_F^2 + \theta_G^2 \, (N_i N_j)^{-\omega} \Big] \, e^{-(x_i - x_j)^2 / 2 \theta_1^2} + \delta_{ij} \big( \sigma_{Q,i}^2 + \theta_2^2 \big) .
    \label{eq:fss_kernel}
\end{equation}
The parameters $p_c$, $\nu$, $b$, $\omega$, and $\theta_{F,G,1,2}$ are obtained by maximizing the log marginal likelihood, or evidence,
\begin{equation}
    \ln \mathcal{L} = -\frac{1}{2} \mathbf{Y}^{T} C^{-1} \mathbf{Y} - \frac{1}{2} \ln \det C - \frac{n}{2} \ln 2\pi ,
    \label{eq:fss_evidence}
\end{equation}
where $\mathbf{Y}$ collects the $n$ centered data points.
The evidence is large when points of different $N$ at nearby $x$ have equal values, so that they fall on a single smooth curve, and no functional form of $F$ or $G$ is assumed.
We maximize $\ln \mathcal{L}$ with a quasi-Newton (L-BFGS-B) algorithm and its analytic gradient, from $20$ starting points.
The critical point $p_c$ is confined to the range of $p$ covered by every size, and $b$ is constrained so that the scaling field remains monotonic in $p$ across the window, which makes $x$ a valid coordinate.

For the PWR2 graph, we also include the next correction to scaling, $N^{-2\omega} H(x)$, where $H$ is a third independent Gaussian process; it adds the term $\theta_H^2 \, (N_i N_j)^{-2\omega}$ inside the square brackets of Eq.~\eqref{eq:fss_kernel}, and $\theta_H$ is fitted together with the other parameters.
Since Eq.~\eqref{eq:fss_model} is recovered for $\theta_H \to 0$, the evidence itself decides whether the second correction is needed.

A single fit depends on two choices that the ansatz does not fix: the window of blocking probabilities around $p_c$ and the smallest system size retained.
Rather than selecting one, we repeat the fit on a grid of windows and minimal sizes, and combine the fits that are mutually consistent; their residual spread provides a systematic error.
The full implementation is available with the code accompanying this work~\cite{code_repo}.

Each fit uses sizes $N \ge n_{\min}$ and points with $|p - p_c| \le h$.
Starting from a guess $p_c^{\rm guess}$, the fit is iterated and the window is re-centered on the fitted $p_c$ until the two agree within $2 \times 10^{-4}$, below the spacing of the $p$ grid.
A configuration $(n_{\min}, h)$ is fitted only if every included size covers the whole window, i.e., has data at both of its edges, no gap larger than $\Delta p = 0.005$ inside it, and at least $8$ points; otherwise it is skipped.
This way, all sizes constrain the same range of $x$, and a size that ends inside the window cannot bias the collapse.
To keep the cost of the Cholesky decomposition of $C$ manageable, at most $100$ evenly spaced points per size enter the evidence; plots of the collapses show all the points instead.
A fit is labeled as failed if any parameter ends on the boundary of its allowed range, except $\theta_2$ at its lower bound, which signals a collapse within the statistical errors, and, in the model with two corrections, one of $\theta_G$ and $\theta_H$ at its lower bound, which leaves a model with a single correction.
The smallest size, $N = 64$, is never included, $n_{\min} \ge 128$, since its deviations from the collapse of the larger sizes exceed what the leading correction can parameterize.

For each graph --- LR and PWR2 --- and value of $\alpha$, we fit every configuration on a grid $n_{\min} \in \{128, 256\}$ and $3$ to $7$ values of $h \in [0.01, 0.1]$, with sizes up to $N = 2048$ ($N = 4096$ for the LR graph at $d_s = 10/3$); $h$ is smaller at small $\alpha$, where $p_c$ approaches $1$.
The statistical errors of all configurations are obtained from a single joint parametric bootstrap.
In each of $100$ replicas $b$, every data point is perturbed once, $Q_i^{(b)} = Q_i + \sigma_{Q,i} \, \eta_i^{(b)}$ with independent standard normal $\eta_i^{(b)}$, and every configuration is refitted to the same replica with the same estimator, with the window held at the center of its central fit.
Distinct $(N, p)$ points are obtained from independent realizations, and the error of Eq.~\eqref{eq:Q_error} agrees with a nonparametric resampling of the realizations within $5\%$.

Fits with different $n_{\min}$ at the same $h$ test whether the remaining size dependence is captured by the corrections of the ansatz.
At each $h$, the differences $\mathbf{D}$ between the values of $\nu$ at the larger $n_{\min}$ and at the smallest one are tested against zero with $\chi^2 = \mathbf{D}^{T} \, \mathrm{Cov}(\mathbf{D})^{-1} \, \mathbf{D}$, where $\mathrm{Cov}(\mathbf{D})$ is estimated from the replicas.
If the corresponding $p$-value is below $0.05$, the smallest $n_{\min}$ is dropped and the test repeated; at least two consistent fits are required, otherwise no configuration at that $h$ is accepted.
For $n_{\min} \in \{128, 256\}$ the test reduces to accepting both fits at a given $h$ if their difference is compatible with zero, and neither otherwise.
The covariance of the differences, rather than the individual error bars, is essential: fits that share the data of the larger sizes are positively correlated, so that their difference has a much smaller error than the individual errors suggest.

The final value of $\nu$ is the mean $\bar{\nu}$ of the $M$ accepted configurations at all $h$, and its statistical error $\sigma_{\rm stat}$ is the standard deviation over the replicas of the mean of the same configurations.
Since accepted configurations are compatible by construction, most of their spread is statistical.
We therefore define the systematic error as the spread in excess of that expected from noise alone,
\begin{equation}
    \sigma_{\rm syst} = \sqrt{\max\big(0, \, s^2 - s_0^2\big)} ,
    \label{eq:syst}
\end{equation}
where $s^2$ is the sample variance of the $M$ central values and $s_0^2$ the average over the replicas of the sample variance of their fluctuations $\nu_k^{(b)} - \langle \nu_k \rangle_b$ around their replica means.
Adding $s$ itself to $\sigma_{\rm stat}$ would count the statistical error twice and make the systematic error grow with the number of configurations.
The total error is $\sigma_{\rm tot} = (\sigma_{\rm stat}^2 + \sigma_{\rm syst}^2)^{1/2}$.
The analysis for $p_c$ is analogous.
The results are collected in Table~\ref{tab:fss_results}, and the full analysis of every $\alpha$, i.e., the data, a representative collapse, and the dependence of $\nu$, $p_c$, and $\omega$ on $h$ and $n_{\min}$, is shown in Figs.~\ref{fig:fss_lr} and~\ref{fig:fss_pwr2}.

The analysis is restricted to $d_s \le 10/3$, i.e., $\alpha \ge 1.6$ for the LR graph and $\alpha \ge 0.6$ for the PWR2 graph.
For smaller $\alpha$, increasingly large system sizes $N$ are required due to stronger finite-size effects, which are beyond what is numerically accessible with our current resources.

\section{Numerical computation of the code distance}
\label{app:qec}
The code distances introduced in Sec.~\ref{sec:qec} are computed directly on sampled interaction graphs.
For the CSS choice NB $= \mathrm{CNOT}$ and B $= \id$, a blocking gate is the absence of a bond and each realization is fully specified by the graph of the non-blocking bonds.
We arrange $N$ qubits on a ring with periodic boundary conditions and for every site $i$ and distance $r = 1, \dots, N/4$ we realize the bond $(i, i+r)$, with site labels understood modulo $N$, independently with probability
\begin{equation}
    \overline{p}_r = (1-p) \, r^{-\alpha} ,
    \label{eq:css_bond_prob}
\end{equation}
i.e., the probability that the pair is connected and hosts an NB gate.
At large distances, where $\overline{p}_r$ is small, we draw the number of bonds at distance $r$ from a binomial distribution and place them uniformly at random; this is equivalent to independent draws for every site, but avoids generating $N$ random numbers per distance.

Both $d_Z$ and $d_X$ are computed on the ring cut open at the meridian between the sites $N$ and $1$.
The bonds that do not cross the meridian form the bulk graph, while each \textit{crossing bond} $(u, v)$ connects a site $u \in \{3N/4+1, \dots, N\}$ to a site $v \in \{1, \dots, N/4\}$, as $r \le N/4$.
We denote by $V_+$ and $V_-$ the sets of the endpoints of the crossing bonds on the two sides of the meridian.

For the systole $d_Z$, we restrict to the winding cycles that cross the meridian exactly once.
Each of them consists of a crossing bond $(u, v)$ closed by a path from $v$ to $u$ in the bulk graph, so that
\begin{equation}
    d_Z = \min_{(u,v)} \big[ 1 + \mathrm{dist}(v, u) \big] ,
    \label{eq:dZ_numerics}
\end{equation}
where the minimum runs over the crossing bonds and $\mathrm{dist}$ is the number of bonds of the shortest bulk path, obtained by breadth-first search from the crossing-bond endpoints.
For a simple cycle the number of bonds equals the number of sites, so that Eq.~\eqref{eq:dZ_numerics} is the weight of $\bar{Z}$.
If no crossing bond has both endpoints in the same connected component of the bulk graph, no winding cycle crosses the meridian once: we set $d_Z = \infty$ and $\mathcal{D} = 0$, as for a realization that encodes no logical qubit.
Equation~\eqref{eq:dZ_numerics} is an upper bound on the systole, since winding cycles that cross the meridian three or more times are not considered.
Comparing with the exact systole for $N \le 2^{15}$, we find that this changes $\langle d_Z \rangle$ by at most $1\%$, and $\langle \mathcal{D} \rangle$ by a few percent, within the statistical errors.

The cosystole $d_X$ is the minimum number of sites whose removal disconnects $V_-$ from $V_+$ in the bulk graph; any such set severs every winding cycle.
We compute it as a maximum flow on an auxiliary directed graph.
Each site $i$ is replaced by two nodes, $i_{\rm in}$ and $i_{\rm out}$, joined by a directed edge $i_{\rm in} \to i_{\rm out}$ of capacity one, and each bulk bond $(i, j)$ by the two directed edges $i_{\rm out} \to j_{\rm in}$ and $j_{\rm out} \to i_{\rm in}$ of infinite capacity.
A source is joined to every $v_{\rm in}$ with $v \in V_-$, and every $u_{\rm out}$ with $u \in V_+$ to a sink, again with infinite capacity.
The capacity of a directed edge is the largest flow it can carry, and by the max-flow min-cut theorem~\cite{ford1956maximal} the maximum flow from the source to the sink equals the minimum total capacity of a set of edges whose removal disconnects the two.
Since only the edges $i_{\rm in} \to i_{\rm out}$ have finite capacity, such a minimal set consists of these edges alone, each corresponding to the removal of one site at unit cost, and the maximum flow equals $d_X$.
The cut separates every site of $V_-$ from every site of $V_+$, including endpoints of different crossing bonds, so that $d_X$ can be nonzero in realizations without a winding cycle; these have $\mathcal{D} = 0$ regardless.

For every realization, we first determine the connected components of the bulk graph.
This decides whether a winding cycle exists and whether $V_-$ and $V_+$ are connected; when they are not, the corresponding search or maximum flow is skipped, which makes the localized phase inexpensive.
All graph computations use the sparse-graph routines of SciPy~\cite{virtanen2020scipy}.
Working on the graph rather than on the tableau gives access to $N = 2^{20} \approx 10^6$ sites, compared with $N \le 4096$ for the tableau simulations of App.~\ref{app:details_cliff}.

The code distance $\mathcal{D} = \min(d_X, d_Z)$ is evaluated for each realization before averaging.
Averages $\langle \cdot \rangle$ are taken over the realizations, with the standard error of the mean as error bar.
The average $\langle d_Z \rangle$ includes only the realizations with a winding cycle, and is shown only where these are at least $99\%$ of the total.

\section{Fidelity estimates for an implementation with reconfigurable neutral-atom arrays}
\label{app:fidelityestimates}
\begin{figure}[t]
    \centering
    \includegraphics{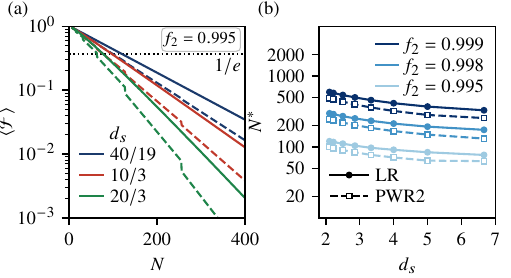}
    \caption{\textbf{Fidelity of one Floquet period.}
    (a) Average fidelity $\langle \mathcal{F} \rangle$ of one period, Eq.~\eqref{eq:mean_fidelity} with $t = 1$, versus system size, for the LR (solid) and PWR2 (dashed) graphs at three spectral dimensions and two-qubit gate fidelity $f_2 = 0.995$.
    The steps of the PWR2 curves occur when a new distance $2^n$ enters the range $r \le N/4$.
    (b) Largest system size $N^*$ with $\langle \mathcal{F} \rangle \ge 1/e$ versus spectral dimension, for three values of $f_2$, at the spectral dimensions of Fig.~\ref{fig:phase_diagram}(b).}
    \label{fig:fidelity}
\end{figure}
Programmable neutral-atom processors combine high-fidelity parallel entangling gates with fast atom rearrangement in optical tweezers~\cite{bluvstein2022quantum,evered2023high,bluvstein2024logical}, which makes them a natural platform for implementing the Floquet circuits considered in this work.
Each Floquet period consists of single-qubit Clifford rotations, applied globally or locally with microwave or Raman pulses; layers of two-qubit Rydberg gates, executed in parallel on pairs of atoms brought within the blockade radius while different pairs remain far apart~\cite{xu2021fast,fromonteil2024hamilton}; and atom rearrangements between successive layers, which change the pairing pattern.
The B gate is a CZ, i.e., a single native Rydberg gate, whereas the NB gate belongs to the iSWAP class (App.~\ref{app:BNB_gates}) and requires two CZ gates combined with single-qubit rotations.

We estimate the fidelity of one Floquet period by counting the two-qubit gates, which dominate the error budget, assuming that they fail independently and neglecting the errors of single-qubit rotations, atom transport, and idling.
We denote by $f_2$ the fidelity of one B or NB gate; since a bond hosts a B gate with probability $p$ and an NB gate otherwise, $f_2 \approx f_{\rm CZ}^{\,2-p}$ to first order in $1 - f_{\rm CZ}$.
Every realized bond hosts one gate, so that the number $G$ of two-qubit gates per period is a sum of independent Bernoulli variables, with the bond probabilities $p_r = r^{-\alpha}$ and the range $r \le N/4$ of App.~\ref{app:details_cliff}.
Its mean and variance,
\begin{equation}
    \langle G \rangle = N \sum_r p_r \equiv c N ,
    \qquad
    \sigma_G^2 = N \sum_r p_r \left( 1 - p_r \right) ,
    \label{eq:gate_count}
\end{equation}
both grow linearly with $N$, so that $G$ is self-averaging, $\sigma_G / \langle G \rangle \sim N^{-1/2}$.
Within the transition windows, the number $c$ of gates per site converges as $N \to \infty$, to $c = \zeta(\alpha)$ for the LR graph and to $c = (1 - 2^{-\alpha})^{-1}$ for the PWR2 graph.
At matched spectral dimension, $\alpha_{\rm PWR2} = \alpha_{\rm LR} - 1$, the PWR2 graph requires $20$--$40\%$ more gates per site than the LR graph, e.g., $c = 2.94$ versus $2.29$ at $d_s = 10/3$, but only $\log_2(N/4) + 1$ distinct distances instead of $N/4$.
Since disjoint gates are executed in parallel, the number of layers is set by the largest number of bonds of a single site, which grows much more slowly than $G$.

The fidelity of one period is then $\mathcal{F} = f_2^{\,G}$, and its average over realizations follows from the generating function of $G$.
Since the graph is the same at every period, after $t$ periods
\begin{equation}
    \langle \mathcal{F}_t \rangle = \prod_r \left[ 1 - \left( 1 - f_2^{\,t} \right) p_r \right]^N \simeq e^{-(1 - f_2)\, c N t} ,
    \label{eq:mean_fidelity}
\end{equation}
where the product runs over the allowed distances.
For a single period, $\langle \mathcal{F} \rangle \ge 1/e$ up to $N^* \simeq 1/[(1 - f_2) c]$, while for $t$ periods the budget is shared between system size and evolution time, $N t \lesssim N^*$.

Figure~\ref{fig:fidelity}(a) shows $\langle \mathcal{F} \rangle$ for one period at $f_2 = 0.995$, comparable to the CZ fidelity of current neutral-atom processors~\cite{evered2023high}, and Fig.~\ref{fig:fidelity}(b) the resulting $N^*$ at the spectral dimensions of Fig.~\ref{fig:phase_diagram}(b).
For $f_2 = 0.995$, $N^*$ ranges from $60$ to $120$, so that a Floquet period can be implemented with $\langle \mathcal{F} \rangle \ge 1/e$ for $N \lesssim 10^2$; $N^*$ grows to several hundred for $f_2 = 0.999$.
Reaching the steady state requires several periods, which reduces the accessible sizes accordingly.

\section{Derivation of the annealed spectral dimension}
\label{app:ds}
The connection probabilities $p_{i,j}$ define the annealed (disorder-averaged) adjacency matrix and the associated graph Laplacian
\begin{equation}
    \mathbb{A}_{ij} = p_{i,j},
    \qquad
    \mathbb{L} = \mathbb{D} - \mathbb{A},
    \qquad
    \mathbb{D}_{ii} = \sum_{j} p_{i,j},
    \label{eq:laplacian-def}
\end{equation}
with eigenpairs $\mathbb{L}\,u_k = \lambda_k u_k$, $0 = \lambda_0 < \lambda_1 \le \lambda_2 \le \dots$, and $u_0 = \mathbf{1}/\sqrt{N}$.
The eigenstates $u_k$ form a basis in which the problem diagonalizes; for a graph with translation symmetry the $u_k$ are Fourier modes.

The spectral dimension $d_s$ is defined through the low-lying spectrum, either via the spectral gap,
\begin{equation}
    \lambda_1 \sim N^{-2/d_s},
    \label{eq:ds-gap}
\end{equation}
or equivalently, via the low-energy power-law behavior of the density of states,
\begin{align}
    \rho(\lambda) = \sum_{k}\delta(\lambda-\lambda_k)\approx \lambda^{d_s/2-1}\,.
    \label{eq:ds-dos}
\end{align}

\subsection{Long-range graph}
\label{app:ds_longrange}
For the long-range graph the bond probability depends on the lattice distance $r_{i,j}$ as the power law
\begin{align}
    p_{i,j} = r_{i,j}^{-\alpha}\,.
\end{align}
Leveraging the translational invariance of the annealed adjacency matrix $\mathbb{A}_{ij}=p_{i,j}$, it is convenient to introduce its Fourier transform
\begin{align}
    \tilde{p}_k(\alpha) = \sum_{r = 1}^{N}\frac{\cos(kr)}{r^{\alpha}}.\label{eq:p-alpha-k}
\end{align}
In the following, we normalize $\tilde{p}_k(\alpha)$ by $\mathcal{N}_\alpha = \sum_r r^{-\alpha}$, so that $\tilde{p}_0(\alpha) = 1$.
For $\alpha > d$, $\mathcal{N}_\alpha$ converges as $N \to \infty$ and the normalization leaves $d_s$ unchanged, while for $\alpha < d$ it is the Kac normalization discussed in App.~\ref{subapp:strong-long-range-p}.
Assuming periodic boundary conditions, we have the usual restriction on the momentum $k\equiv k_n = 2\pi n/N$ with $n\in \mathbb{Z}$ and $n = -\lceil N/2\rceil+1,\dots,\lfloor N/2\rfloor$ (the lattice spacing has been set to 1). Since $\tilde{p}_0(\alpha) = \mathbb{D}_{ii}$, the Laplacian eigenvalues are $\lambda_k = \tilde{p}_0(\alpha)-\tilde{p}_k(\alpha)$, and the spectral gap, setting the spectral dimension, is determined by the minimum gap with respect to the zero mode
\begin{align}
\delta\lambda =\min_{k\neq 0}[\tilde{p}_0(\alpha)-\tilde{p}_k(\alpha)]\approx N^{-2/d_s}\,.
\end{align}
If the minimum is attained by the second lowest lying Fourier mode $k_1 = 2\pi/N$ and the $N\to\infty$ limit corresponds to a well-defined continuous limit in $k$-space, it follows that the spectral dimension can also be determined by the dispersion relation close to $k\approx 0$ using the relation
\begin{align}
\tilde{p}_0(\alpha)-\tilde{p}_k(\alpha)\approx |k|^{2d/d_s}\,.\label{eq:p-dispersion-ds}
\end{align}

\subsubsection{Weak long-range (\texorpdfstring{$\alpha>d$}{alpha > d})}
In the weak long-range regime $\alpha>d$ the $N\to\infty$ limit corresponds to a continuum limit in the Fourier modes, allowing us to substitute the discrete values $k_n$ with the continuous variable $k\in[-\pi,\pi)$.

For simplicity, we start from the one dimensional case $d = 1$. Then, for $\alpha>1$ and in the $N\to\infty$ limit the Fourier transform of the connection probabilities reads~\cite{defenu2019universal}
\begin{align}
    \tilde{p}_k(\alpha) \approx \frac{1}{2\zeta(\alpha)}\left[\mathrm{Li}_\alpha(e^{ik})+\mathrm{Li}_\alpha(e^{-ik})\right]\label{eq:p-alpha-weak-LR}
\end{align}
where $\mathrm{Li}_x(z) = \sum_{n = 1}^{\infty}z^n/n^x$ is the polylogarithm and $\zeta(x)$ is the Riemann zeta function~\cite{abramowitz1964handbook}.
In order to determine the spectral dimension $d_s$, we are interested in the low $k$ modes, which determine the dispersion relation. This is obtained by taking the Taylor expansion of Eq.~\eqref{eq:p-alpha-weak-LR} around $k = 0$ leading to~\cite{defenu2019universal}
\begin{align}
\tilde{p}_k &= 1+\sin\left(\frac{\alpha\pi}{2}\right)\frac{\Gamma(1-\alpha)}{\zeta(\alpha)}|k|^{\alpha-1}+\mathcal{O}(|k|^2),
\end{align}
for $1<\alpha<3$,
\begin{align}
\tilde{p}_k &=  1+\frac{2\ln|k|-3}{4\zeta(3)}|k|^2 + \mathcal{O}(|k|^3),
\end{align}
for $\alpha = 3$,
\begin{align}
\tilde{p}_k &=  1-\frac{\zeta(\alpha-2)}{2\zeta(\alpha)}|k|^2 + \mathcal{O}(|k|^{\alpha-1}),
\end{align}
for $\alpha > 3$.
In order to generalize the above results to the $d>1$ case we follow the standard procedure for Fourier transforming a radial function: we switch to spherical coordinates and integrate over all the angles, obtaining~\cite{defenu2024outofequilibrium}
\begin{align}
    \tilde{p}_{\boldsymbol{k}}\approx \frac{2^{d/2-1}\Gamma(d/2)\int_{1}^{\infty} d\rho\rho^{d-1-\alpha}\mathcal{J}_{d/2-1}(|\boldsymbol{k}|\rho)(|\boldsymbol{k}|\rho)^{1-d/2}}{\int_1^\infty d\rho\rho^{d-1-\alpha}},\label{eq:p-Fourier-d}
\end{align}
where $\mathcal{J}_\nu(x)$ is the standard Bessel function of order $\nu$~\cite{abramowitz1964handbook}. Expanding Eq.~\eqref{eq:p-Fourier-d} in the $|\boldsymbol{k}|\approx 0$ limit we obtain
\begin{align}
\tilde{p}_{\boldsymbol{k}}&\approx 1-A(\alpha)|\boldsymbol{k}|^{\alpha-d}+B(\alpha)|\boldsymbol{k}|^2+\mathcal{O}(|\boldsymbol{k}|^4)
\end{align}
where
\begin{align}
    A(\alpha) = \frac{\Gamma(d/2)\Gamma((2-\alpha+d)/2)}{2^{\alpha-d}\Gamma(\alpha/2)}\,,
    \end{align}
and
\begin{align}
    B(\alpha) = \frac{\alpha-d}{2d^2+4d-2\alpha d}\,.
\end{align}
The first (second) term governs the asymptotic low-momentum behavior of $\tilde{p}_{\boldsymbol{k}}$ for $d<\alpha<d+2$ (respectively $\alpha>d + 2$), leading to the dispersion relation
\begin{align}
   \tilde{p}_{\boldsymbol{0}}(\alpha)-\tilde{p}_{\boldsymbol{k}}(\alpha)\approx\begin{cases}
       |\boldsymbol{k}|^{\alpha-d} &d<\alpha<d+2\\
       |\boldsymbol{k}|^{2} &\alpha>d+2
   \end{cases}\,.
\end{align}
Finally comparing this result with Eq.~\eqref{eq:p-dispersion-ds} we obtain the spectral dimension for the long-range graph
\begin{align}
    d_s = \begin{cases}
       2d/(\alpha-d)  &d<\alpha<d+2\\
       d              &\alpha>d+2
   \end{cases}\,.
\end{align}

\subsubsection{Strong long-range (\texorpdfstring{$0 < \alpha < d$}{0 < alpha < d})}
\label{subapp:strong-long-range-p}
The situation changes dramatically in the strong long-range regime $\alpha < d$. Here $\mathcal{N}_\alpha$ is a genuine Kac normalization~\cite{kac1963van}, needed to keep the average degree $\mathbb{D}_{ii}$ finite, and its scaling with the number of sites $N\gg1$ depends on $\alpha$ as
\begin{align}
    \mathcal{N}_\alpha\approx \begin{cases}
        N^{1-\alpha/d} &\mathrm{if}\,\,\alpha<d\\
        \ln N &\mathrm{if}\,\,\alpha = d\\
        \mathrm{const.} &\mathrm{if}\,\,\alpha>d
    \end{cases}\,,\label{eq:kac-scaling-p}
\end{align}
where the constant reduces to $\zeta(\alpha)$ for $d = 1$.
As shown by this scaling, $\mathcal{N}_\alpha$ diverges at large $N$ for $\alpha<d$. Accordingly, the thermodynamic limit of Eq.~\eqref{eq:p-alpha-k} must be carefully considered. For simplicity, we show this for $d = 1$, where it is convenient to write Eq.~\eqref{eq:p-alpha-k} explicitly for large $N$ as
\begin{align}
    \lim_{N\to\infty}\frac{1}{\mathcal{N}_\alpha}\sum_{r = 1}^{N/2-1}\frac{\cos(kr)}{r^\alpha}\approx\frac{c_\alpha}{N}\sum_{r = 1}^{N/2}\frac{\cos(2\pi n \frac{r}{N})}{(r/N)^\alpha},
\end{align}
with $c_\alpha = \lim_{N\to\infty} N^{1-\alpha}/\mathcal{N}_\alpha = (1-\alpha)\,2^{1-\alpha}$, which ensures $\tilde{p}_0(\alpha) = 1$.
Due to the $1/N$ scaling of the discrete momenta on the lattice, the summation depends only on the variable $r/N$. Therefore, for $N\to\infty$, we can take the continuum limit by transforming the sum over $r$ into an integral with respect to $s = r/N$, leading to~\cite{defenu2021metastability}
\begin{align}
    \tilde{p}_n(\alpha) = \lim_{N\to\infty}\tilde{p}_k(\alpha) = c_\alpha\int_0^{1/2} ds\frac{\cos(2\pi n s)}{s^\alpha}.\label{eq:p-alpha-spectrum}
\end{align}
As a consequence the spectrum remains discrete even as $N\to\infty$. Specifically, for $\alpha<d$, the gap between neighboring eigenvalues $\lambda_{n+1}-\lambda_n$, labeled by the consecutive momenta $k_n$, $k_{n+1}$ in Eq.~\eqref{eq:p-alpha-k}, does not vanish in the thermodynamic limit, as it would for $\alpha>d$. Therefore, the eigenvalues depend only on the integer index $n\in \mathbb{Z}$ rather than on the continuous momentum $k$:
\begin{align}
    \lambda_n = \tilde{p}_0(\alpha) - \tilde{p}_n(\alpha).\label{eq:p-discrete-spectrum}
\end{align}
Notably, for $\alpha = 0$, we find that $\tilde{p}_n(\alpha)\to\delta_{n,0}$, leading to a fully degenerate discrete spectrum as described by Eq.~\eqref{eq:p-discrete-spectrum}: $\lambda_n = \tilde{p}_0(\alpha)$ for $n\neq 0$ and $\lambda_n = 0$ for $n=0$. Additionally, we observe that the eigenvalues $\lambda_n$ are not densely distributed. Instead, each eigenvalue is isolated, with the only accumulation point occurring at the maximum $\max_n\lambda_n = \tilde{p}_0(\alpha)$. This follows from the Riemann--Lebesgue lemma~\cite{last1996quantum}, which implies
\begin{align}
\lim_{n\to\infty}\tilde{p}_n(\alpha) = 0.
\end{align}
Finally, since the spectral gap $\delta\lambda$ remains finite in the thermodynamic limit, and more precisely equal to
\begin{align}
    \delta\lambda\approx 1-c_\alpha\int_0^{1/2} ds\frac{\cos(2\pi s)}{s^\alpha},
\end{align}
it follows that the spectral dimension of the long-range graph in the strong long-range regime $0<\alpha<d$ is always infinite
\begin{align}
    d_s = \infty\quad\mathrm{if}\quad0<\alpha<d\,.
\end{align}

\subsubsection{Role of an exponential cutoff}\label{app:exp-cutoff-p}
We analyze the effect of introducing an exponential cutoff in the connection probability, replacing the pure power-law profile with
\begin{align}
p_r \sim r^{-\alpha} e^{-\kappa r}.
\end{align}
The Fourier transform then reads
\begin{align}
    \tilde{p}_k(\alpha) = \sum_{r=1}^N\frac{\cos(kr)e^{-\kappa r}}{r^{\alpha}},
\end{align}
and the spectral gap between the zero mode and the first Fourier mode $k_1 = 2\pi/N$ is
\begin{align}
\delta\lambda
= \tilde{p}_0(\alpha) - \tilde{p}_{k_1}(\alpha)
= \sum_{r=1}^{N} \frac{\left[1-\cos(k_1 r)\right] e^{-\kappa r}}{r^{\alpha}}.
\end{align}
In several experimental settings, the exponential cutoff scales with system size as $\kappa = \Lambda/N$, where $\Lambda=\mathcal{O}(1)$~\cite{schuckert2025observation}. Making the $N$ dependence explicit through $k_1=2\pi/N$ and $\kappa=\Lambda/N$, we obtain
\begin{align}
    \delta\lambda =\frac{1}{N^\alpha}\sum_{r=1}^N\frac{[1-\cos(2\pi (r/N))]e^{-\Lambda (r/N)}}{(r/N)^{\alpha}}\,.
\end{align}
The $N\to\infty$ limit then corresponds to a continuum limit with respect to the variable $x=r/N$, allowing us to pass from a discrete sum over $r$ to an integral over $x$
\begin{align}
    \delta\lambda \approx N^{1-\alpha}\int_{1/N}^1dx\frac{[1-\cos(2\pi x)]e^{-\Lambda x}}{x^{\alpha}}\,.
\end{align}
The convergence of the integral as $N\to\infty$ is controlled by the integrand behavior as $x\sim1/N\to0$
\begin{align}
    \frac{[1-\cos(2\pi x)]e^{-\Lambda x}}{x^{\alpha}}\sim x^{2-\alpha}\,.
\end{align}
Similarly to the case without any exponential cutoff, this allows us to identify two regimes: if $0<\alpha<3$ the integral converges in the thermodynamic limit, yielding $\delta\lambda\sim N^{1-\alpha}$; if $\alpha>3$, the integral diverges as $\sim N^{\alpha-3}$, leading to $\delta\lambda\sim N^{-2}$.
Therefore, the presence of an exponential cutoff does not modify the scaling of the spectral gap, provided that the decay rate scales as $\kappa \sim 1/N$ with the system size. Consequently, the spectral dimension remains unchanged with respect to the pure power-law case.

\subsection{PWR2 graph}
\label{app:ds_pwr2}
Here we analyze the low-energy structure of the spectrum of the power-of-two (PWR2) graph. The PWR2 graph is defined on a one-dimensional ring of $N$ sites, with connection probabilities linking only pairs of sites whose distance is an integer power of two. Explicitly, the connection probabilities read
\begin{align}
p_{i,j} =
\begin{cases}
|i-j|^{-\alpha}, & \text{if } |i-j| = 2^{n},\\
0, & \text{otherwise}.
\end{cases}
\end{align}
with $n=0,1,\dots,\log_2(N)-1$.
In this case, the scaling of the Kac normalization reads
\begin{align}
\mathcal{N}_{\alpha,\mathrm{PWR2}} = \sum_{n=0}^{\log_2(N)-1} 2^{-\alpha n}.
\end{align}
This geometric series can be summed explicitly, yielding in the large-$N$ limit three distinct regimes:
\begin{align}
\mathcal{N}_{\alpha,\mathrm{PWR2}} \approx
\begin{cases}
\dfrac{N^{|\alpha|}}{2^{|\alpha|}-1}, & \alpha < 0,\\[6pt]
\log_2(N), & \alpha = 0,\\[6pt]
\dfrac{1}{1-2^{-\alpha}}, & \alpha > 0.
\end{cases}
\end{align}
For $\alpha<0$, the connection probability increases with distance, leading to a divergence of $\mathbb{D}_{ii}$ with system size; we will therefore need to introduce a normalization factor scaling as $N^{\alpha}$ to cure this divergence, ensuring a finite degree and a meaningful thermodynamic limit also in this case.

In the absence of spatial disorder, translational invariance allows us to diagonalize $\mathbb{A}_{ij}=p_{i,j}$ in momentum space. Introducing the discrete Fourier transform, the spectrum reads
\begin{align}
\tilde{p}_k(\alpha)
=
\sum_{n=0}^{\log_2(N)-1}
2^{-\alpha n}\cos\!\left(k\,2^n\right),
\label{eq:p-spectrum-PWR2}
\end{align}
where $k=2\pi m/N$ with $m = -\lceil N/2\rceil+1,\dots,\lfloor N/2\rfloor$, assuming periodic boundary conditions. Also in this case the spectral gap is determined by the minimum gap with respect to the zero mode
\begin{align}
\delta\lambda =\min_{k\neq 0}[\tilde{p}_0(\alpha)-\tilde{p}_k(\alpha)].
\end{align}

\subsubsection{The \texorpdfstring{$\alpha = 0$}{alpha = 0} case}
For $\alpha=0$, all nonzero connection probabilities have equal strength. Equation~\eqref{eq:p-spectrum-PWR2} reduces to a binary sum of cosine functions. The minimal gap is attained at momentum $k=\pi$, for which
\begin{align}
\tilde{p}_0(\alpha) - \tilde{p}_\pi(\alpha)
&=
\sum_{n=0}^{\log_2(N)-1}
\left[1-\cos(2^n\pi)\right]=2,
\end{align}
independently of system size. Thus, similarly to the strong long-range graph with $\alpha<d$ (see App.~\ref{subapp:strong-long-range-p}), the spectrum of the PWR2 graph at $\alpha=0$ remains gapped in the thermodynamic limit~\cite{kuriyattil2025entangled}. As a consequence, the effective spectral dimension diverges, $d_s\to\infty$.

\subsubsection{The \texorpdfstring{$\alpha > 0$}{alpha > 0} case}
For $\alpha>0$, the sum in Eq.~\eqref{eq:p-spectrum-PWR2} converges as $N\to\infty$ and defines a classical Weierstrass function~\cite{weierstrass1988,hardy1916weierstrass},
\begin{align}
W(x) = \sum_{n=0}^{\infty} a^{n}\cos(x b^{n}),
\end{align}
with parameters
\begin{align}
a = 2^{-\alpha}, \qquad b=2,
\end{align}
satisfying $0<a<1$ and $b>1$. For $ab\ge1$, i.e., $\alpha\le1$, the Weierstrass function is continuous everywhere but nowhere differentiable~\cite{hardy1916weierstrass}, and exhibits a self-similar fractal structure.
To extract the effective dispersion at small momenta, we first work at finite $N$ and expand the cosine in Eq.~\eqref{eq:p-spectrum-PWR2} for small $k$:
\begin{align}
    \tilde{p}_0(\alpha)-\tilde{p}_k(\alpha) &\approx\frac{k^2}{2}
\sum_{n=0}^{\log_2(N)-1}
2^{n(2-\alpha)}
\nonumber\\
&=\frac{k^2}{2}\frac{1-N^{2-\alpha}}{1-2^{2-\alpha}}\,.
\end{align}
Then, to obtain the correct power law behavior as $N\to\infty$, one must take into account that the smallest nonzero momentum scales as $k_1=2\pi/N$. Sending $N\to\infty$ while keeping $kN=\mathcal{O}(1)$, two distinct regimes emerge
\begin{align}
\tilde{p}_0(\alpha) - \tilde{p}_k(\alpha) \approx
\begin{cases}
k^2 & \alpha > 2, \\
k^{\alpha} & 0 < \alpha < 2.
\end{cases}\label{eq:p-PWR2-dispersion-app}
\end{align}
This leads to the spectral dimension
\begin{align}
d_s =
\begin{cases}
1 & \alpha > 2, \\
2/\alpha & 0 < \alpha < 2.
\end{cases}
\end{align}
A more formal derivation follows from rigorous bounds on the Weierstrass function~\cite{david2018bypassing} close to its maximum at $x = 0$, where all the cosines are equal to one,
\begin{align}
    C_{\inf}|k|^{2-d_H}\leq W(0)-W(k)\leq C_{\sup}|k|^{2-d_H},
\end{align}
for small $|k|$, where $C_{\inf}$ and $C_{\sup}$ are strictly positive constants and $d_H$ denotes the Hausdorff (fractal) dimension. For the Weierstrass function with $ab>1$, i.e., $0<\alpha<1$, this dimension is known to be related to the $a$ and $b$ parameters as~\cite{hunt1998thehausdorff,david2018bypassing}
\begin{align}
    d_H = 2+\frac{\ln a}{\ln b} = 2-\alpha,\label{eq:p-PWR2-Hausdorff-dimension}
\end{align}
where in the last equality we used the specific parameters of the PWR2 graph. Since $W(0)-W(k)=\tilde{p}_0(\alpha)-\tilde{p}_k(\alpha)$ as $N\to\infty$, taking $k\to 0$ immediately reproduces the dispersion scaling in Eq.~\eqref{eq:p-PWR2-dispersion-app}. This also yields a direct relation between spectral and fractal dimensions
\begin{align}
    d_s = 2(2-d_H)^{-1},
\end{align}
which provides a general framework for defining the spectral dimension of graph spectra generated by Weierstrass functions with arbitrary $a$ and $b$.

\subsubsection{The \texorpdfstring{$\alpha < 0$}{alpha < 0} case}
For $\alpha<0$, extensivity requires the introduction of a normalization factor $N^{\alpha}$, leading to the rescaled spectrum
\begin{align}
\tilde{p}_k(\alpha)
&=
\sum_{n=0}^{\log_2(N)-1}
N^\alpha2^{-\alpha n}\cos\left(k\,2^n\right)\notag\\
&=\sum_{n=0}^{\log_2(N)-1} 2^{\alpha\left[\log_2(N)-n\right]}\cos\left(2^n k\right).
\label{eq:p-spectrum-PWR2-negative-alpha}
\end{align}
Because $\alpha<0$, the dominant contributions now come from large distances $n\sim\log_2(N)$ rather than from short-range terms. This becomes explicit upon performing the change of variables $n'=\log_2(N)-1-n$ (related to the so-called Monna map~\cite{monna1952transformation}), yielding
\begin{align}
\tilde{p}_k(\alpha) &= 2^{\alpha}\sum_{n=0}^{\log_2(N)-1} 2^{\alpha n}\cos\left(2^{-n}\tilde{k}\right)\,,
\end{align}
where we introduced the rescaled momenta $\tilde{k}=Nk/2$. Now recalling that $k=k_m = 2\pi m/N$, with $m$ an integer such that $m = -N/2+1,\dots,N/2$ (assuming periodic boundary conditions), we notice that the first mode after the zero mode $k_0 = 0$ is no longer given by $k_1= 2\pi/N$. Indeed, in the sum for $m=1$ we have that $\tilde{k}_1 = Nk_1/2 = \pi$, meaning that this mode remains gapped in the thermodynamic limit $N\to\infty$. Instead, similarly to the $\alpha = 0$ case, the minimal gap corresponds to the $k=\pi$ mode. The corresponding spectral gap is given by
\begin{align}
    \delta\lambda &= \tilde{p}_0(\alpha)-\tilde{p}_\pi(\alpha) \\&=N^\alpha\sum_{n=0}^{\log_2(N)-1}2^{-\alpha n}[1-\cos\left(\pi\,2^n\right)]\,.
\end{align}
Since $\cos(\pi 2^n)=1$ for all $n\geq1$, only the $n=0$ term contributes, yielding
\begin{align}
\delta\lambda = 2N^{\alpha} = 2N^{-|\alpha|}.
\end{align}
This scaling implies a spectral dimension
\begin{align}
d_s = \frac{2}{|\alpha|}.
\end{align}

\end{appendix}

\else
\fi

\bibliography{references}

\end{document}